\documentclass[journal,draftclsnofoot,onecolumn,12pt,oneside]{IEEEtran} 

\usepackage{cite}

\ifCLASSINFOpdf
  \usepackage[pdftex]{graphicx}
  \usepackage{epstopdf} 
\else
\fi
\usepackage{amsmath}
\usepackage{algorithm}
\usepackage{algorithmic}

  \usepackage[caption=false,font=footnotesize]{subfig}
\usepackage{url}

\usepackage{amsthm}

\newtheoremstyle{customtheorem}{0pt}{0pt}{\itshape}{12pt}{\bfseries}{.}{0.5em}{}
\newtheoremstyle{customdefinition}{0pt}{0pt}{}{12pt}{\bfseries}{.}{0.5em}{}

\theoremstyle{plain}
\newtheorem{theorem}{Theorem}

\theoremstyle{plain}
\newtheorem{corollary}{Corollary}[theorem]

\theoremstyle{plain}

\theoremstyle{definition}

\theoremstyle{remark}
\newtheorem{remark}{Remark}

\newcommand{\comment}[1]{{}}

\usepackage{amssymb}
\usepackage{xspace}
\usepackage{bm}

\newcommand{\set}[1]{\ensuremath{\mathcal{#1}}\xspace} 
\newcommand{\mat}[1]{\ensuremath{\mathbf{#1}}\xspace} 
\renewcommand{\vec}[1]{\ensuremath{\mathbf{#1}}\xspace} 

\newcommand{\parens}[1]{\ensuremath{\left(#1\right)}\xspace}
\newcommand{\brackets}[1]{\ensuremath{\left[#1\right]}\xspace}
\newcommand{\braces}[1]{\ensuremath{\left\{#1\right\}}\xspace}
\newcommand{\bars}[1]{\ensuremath{\left\vert#1\right\vert}\xspace}
\newcommand{\doublebars}[1]{\ensuremath{\left\Vert#1\right\Vert}\xspace}

\newcommand{\complex}{\ensuremath{\mathbb{C}}\xspace}

\renewcommand{\j}{\ensuremath{\mathrm{j}}}

\newcommand{\card}[1]{\bars{#1}}

\newcommand{\setcomplex}{\ensuremath{\complex}}

\newcommand{\setvector}[2]{\ensuremath{#1^{#2 \times 1}}\xspace}

\newcommand{\setvectorcomplex}[1]{\setvector{\setcomplex}{#1}}

\newcommand{\setmatrix}[3]{\ensuremath{#1^{#2 \times #3}}\xspace}

\newcommand{\setmatrixcomplex}[2]{\setmatrix{\setcomplex}{#1}{#2}}

\newcommand{\complexpow}[2]{\ensuremath{\complex^{#1 \times #2}}\xspace}

\newcommand{\onevec}[1]{\ensuremath{\vec{1}_{#1}}\xspace}

\newcommand{\eyemat}[1]{\ensuremath{\mat{I}_{#1}}\xspace}

\newcommand{\inv}{\ensuremath{^{-1}}\xspace}

\newcommand{\trans}{\ensuremath{^{\mathrm{T}}}\xspace}
\newcommand{\ctrans}{\ensuremath{^{{*}}}\xspace}

\newcommand{\had}{\ensuremath{\odot}\xspace}
\newcommand{\entry}[2]{\ensuremath{\brackets{#1}_{#2}}\xspace}

\newcommand{\logtwodet}[1]{\ensuremath{\mathrm{log}_{2}\bars{#1}}}

\newcommand{\diag}[1]{\ensuremath{\mathrm{diag}\parens{#1}}}

\newcommand{\abs}[1]{\ensuremath{\mathrm{abs}\parens{#1}}}

\newcommand{\svmax}[1]{\ensuremath{\sigma_{\mathrm{max}}\parens{#1}}\xspace}

\newcommand{\svmaxsq}[1]{\ensuremath{\sigma_{\mathrm{max}}^2\parens{#1}}\xspace}

\newcommand{\pnorm}[2]{\ensuremath{\doublebars{#2}_{#1}}\xspace}

\newcommand{\normfro}[1]{\pnorm{\mathrm{F}}{#1}}

\newcommand{\svdop}[1]{\ensuremath{\mathrm{SVD}\parens{#1}}}

\newcommand{\frobeniustwo}[1]{\normfro{#1}^2}

\newcommand{\distcgauss}[2]{\ensuremath{\mathcal{N}_{\complex}\parens{#1,#2}}\xspace} 
\newcommand{\distuniform}[2]{\ensuremath{\mathrm{Unif}\parens{#1,#2}}\xspace}

\newcommand{\ev}[1]{\ensuremath{\mathbb{E}\brackets{#1}}\xspace}

\newcommand{\cgauss}[2]{\distcgauss{#1}{#2}} 

\newcommand{\maxkop}[1]{\ensuremath{\mathrm{maxk}\parens{#1}}\xspace}

\newcommand{\st}{\ensuremath{\mathrm{s.t.~}}\xspace}

\newcommand{\ispsd}{\ensuremath{\succeq}}

\newcommand{\node}[1]{\unskip\ensuremath{^{^{\left(#1\right)}}}\xspace}
\newcommand{\ctransnode}[1]{\ensuremath{^{^{\left(#1\right)}{*}}}\xspace}

\newcommand{\symboltime}{\ensuremath{T}}

\newcommand{\quantize}[2]{\ensuremath{\mathcal{Q}\parens{#2}}\xspace}

\newcommand{\Ptx}{\ensuremath{P_{\mathrm{tx}}\xspace}}
\newcommand{\Ptxwatts}{\ensuremath{\tilde{P}_{\mathrm{tx}}\xspace}}

\newcommand{\powertotmaxlna}{\ensuremath{P_{\mathrm{LNA}}^{\mathrm{max}}}\xspace}

\newcommand{\powersimaxlnawatts}{\ensuremath{\tilde{P}_{\mathrm{SI,LNA}}^{\mathrm{max}}}\xspace}

\newcommand{\powersimaxadcwatts}{\ensuremath{\tilde{P}_{\mathrm{SI,ADC}}^{\mathrm{max}}}\xspace}

\newcommand{\Pt}{\Ptx}

\newcommand{\snr}{\ensuremath{\mathrm{SNR}}\xspace}

\newcommand{\rxresponsevector}{\ensuremath{\vec{a}_{\mathrm{rx}}}\xspace}
\newcommand{\txresponsevector}{\ensuremath{\vec{a}_{\mathrm{tx}}}\xspace}

\newcommand{\aoa}{\ensuremath{\mathrm{AoA}}\xspace}
\newcommand{\aod}{\ensuremath{\mathrm{AoD}}\xspace}

\newcommand{\numrays}{\ensuremath{N_{\mathrm{rays}}}\xspace}

\newcommand{\Nt}{\ensuremath{N_\mathrm{t}}\xspace} 
\newcommand{\Nr}{\ensuremath{N_\mathrm{r}}\xspace} 

\newcommand{\Lt}{\ensuremath{L_{\mathrm{t}}}\xspace} 
\newcommand{\Lr}{\ensuremath{L_{\mathrm{r}}}\xspace} 

\newcommand{\symbvec}{\ensuremath{\vec{s}}\xspace}

\newcommand{\symbvecest}{\ensuremath{\hat{\vec{s}}}\xspace}

\newcommand{\Ns}{\ensuremath{N_\mathrm{s}}\xspace} 

\newcommand{\channel}{\ensuremath{\mat{H}}\xspace}

\newcommand{\pre}{\ensuremath{\mat{F}}\xspace}
\newcommand{\prebb}{\ensuremath{\pre_{\mathrm{BB}}}\xspace}
\newcommand{\prerf}{\ensuremath{\pre_{\mathrm{RF}}}\xspace}

\newcommand{\prerftr}{\ensuremath{\mat{F}_{\mathrm{tr}}}\xspace} 

\newcommand{\prerfcb}{\ensuremath{\mathcal{F}_{\mathrm{RF}}}\xspace}

\newcommand{\com}{\ensuremath{\mat{W}}\xspace}
\newcommand{\combb}{\ensuremath{\com_{\mathrm{BB}}}\xspace}
\newcommand{\comrf}{\ensuremath{\com_{\mathrm{RF}}}\xspace}

\newcommand{\comrftr}{\ensuremath{\com_{\mathrm{tr}}}\xspace}

\newcommand{\comrfcb}{\ensuremath{\mathcal{W}_{\mathrm{RF}}}\xspace}

\newcommand{\xbb}{\ensuremath{\mat{X}_{\mathrm{BB}}}\xspace} 

\newcommand{\noisestd}{\ensuremath{\sigma_{\mathrm{n}}}\xspace}
\newcommand{\noisevar}{\ensuremath{\noisestd^2}\xspace}

\newcommand{\noisevec}{\ensuremath{\vec{n}}\xspace}

\newcommand{\lnagain}{\ensuremath{{G}_{\mathrm{rx}}}\xspace}

\newcommand{\lnafun}[1]{\ensuremath{\mathcal{L}\parens{#1}}}
\newcommand{\lnafunlin}[1]{\ensuremath{\mathcal{L}_{\mathrm{lin}}\parens{#1}}}
\newcommand{\lnafunsat}[1]{\ensuremath{\mathcal{L}_{\mathrm{sat}}\parens{#1}}}

\newcommand{\ylna}{\ensuremath{\vec{y}_{\mathrm{LNA}}}\xspace}

\newcommand{\ylnadot}{\ensuremath{\vec{y}_{(\cdot),\mathrm{LNA}}}\xspace}

\newcommand{\ylnades}{\ensuremath{\vec{y}_{\mathrm{des,LNA}}}\xspace}
\newcommand{\ylnaint}{\ensuremath{\vec{y}_{\mathrm{int,LNA}}}\xspace}
\newcommand{\ylnanoise}{\ensuremath{\vec{y}_{\mathrm{noise,LNA}}}\xspace}

\newcommand{\etalna}{\ensuremath{\eta_{\mathrm{LNA}}}\xspace}

\newcommand{\adcfun}[2]{\ensuremath{\quantize{#1}{#2}}}
\newcommand{\adcbits}{\ensuremath{b}\xspace}

\newcommand{\yadc}{\ensuremath{\vec{y}_{\mathrm{ADC}}}\xspace}

\newcommand{\yadcdot}{\ensuremath{\vec{y}_{(\cdot),\mathrm{ADC}}}\xspace}

\newcommand{\yadcdes}{\ensuremath{\vec{y}_{\mathrm{des,ADC}}}\xspace}
\newcommand{\yadcint}{\ensuremath{\vec{y}_{\mathrm{int,ADC}}}\xspace}
\newcommand{\yadcnoise}{\ensuremath{\vec{y}_{\mathrm{noise,ADC}}}\xspace}

\newcommand{\etaadc}{\ensuremath{\eta_{\mathrm{ADC}}}\xspace}

\newcommand{\equant}{\ensuremath{\ve_{\mathrm{quant}}}\xspace}

\newcommand{\Mt}{\ensuremath{M_{\mathrm{t}}}\xspace}
\newcommand{\Mr}{\ensuremath{M_{\mathrm{r}}}\xspace}

\newcommand{\ydig}{\ensuremath{\vy_{\mathrm{dig}}}}

\newcommand{\Mtx}{\ensuremath{\mathcal{M}_{\mathrm{tx}}}\xspace}
\newcommand{\Mrx}{\ensuremath{\mathcal{M}_{\mathrm{rx}}}\xspace}

\newcommand{\Jtx}{\ensuremath{\mathcal{J}_{\mathrm{tx}}}\xspace}
\newcommand{\Jrx}{\ensuremath{\mathcal{J}_{\mathrm{rx}}}\xspace}

\newcommand{\Tij}{\ensuremath{\mathcal{T}_{ij}}\xspace}
\newcommand{\Tki}{\ensuremath{\mathcal{T}_{ki}}\xspace}

\newcommand{\Rij}{\ensuremath{\mathcal{R}_{ij}}\xspace}
\newcommand{\Rki}{\ensuremath{\mathcal{R}_{ki}}\xspace}

\newcommand{\Cij}{\ensuremath{\mathcal{C}_{ij}}\xspace}
\newcommand{\Cki}{\ensuremath{\mathcal{C}_{ki}}\xspace}

\newcommand{\Iij}{\ensuremath{\mathcal{I}_{ij}}\xspace}

\newcommand{\Kij}{\ensuremath{K_{ij}}\xspace}
\newcommand{\Kki}{\ensuremath{K_{ki}}\xspace}

\def\va{{\vec{a}}}
\def\vb{{\vec{b}}}

\def\ve{{\vec{e}}}

\def\vx{{\vec{x}}}
\def\vy{{\vec{y}}}

\def\mA{{\mat{A}}}
\def\mB{{\mat{B}}}

\def\mH{{\mat{H}}}
\def\mI{{\mat{I}}}

\def\mM{{\mat{M}}}

\def\mP{{\mat{P}}}
\def\mQ{{\mat{Q}}}
\def\mR{{\mat{R}}}

\def\mU{{\mat{U}}}
\def\mV{{\mat{V}}}

\def\mSigma{{\mat{\Sigma}}}

\usepackage{xspace}
\usepackage[acronym,nogroupskip,nonumberlist,nopostdot]{glossaries}
\makeglossaries

\usepackage{xcolor}

\newacronym{snr}{SNR}{signal-to-noise ratio}
\newacronym{sinr}{SINR}{signal-to-interference-plus-noise ratio}
\newacronym{inr}{INR}{interference-to-noise ratio}
\newacronym{sir}{SIR}{signal-to-interference ratio}
\newacronym{sqr}{SQR}{signal-to-quantization-noise ratio}
\newacronym{sqnr}{SQNR}{signal-to-quantization-plus-noise ratio}
\newacronym{ian}{IAN}{interference as noise}
\newacronym{ber}{BER}{bit error rate}
\newacronym{pn}{PN}{pseudorandom noise}
\newacronym{bfsk}{BFSK}{binary frequency shift keying}
\newacronym{fh}{FH}{frequency-hopped}
\newacronym{fh-bfsk}{FH-BFSK}{frequency-hopped binary frequency shift keying}
\newacronym{crc}{CRC}{cyclic redundancy check}
\newacronym{isi}{ISI}{intersymbol interference}
\newacronym{dsss}{DSSS}{direct-sequence spread spectrum}
\newacronym{ofdm}{OFDM}{orthogonal frequency-division multiplexing}
\newacronym{ofdma}{OFDMA}{orthogonal frequency-division multiple access}
\newacronym{sdr}{SDR}{software-defined radio}
\newacronym{tx}{TX}{transmitter}
\newacronym{rx}{RX}{receiver}
\newacronym{fdd}{FDD}{frequency-division duplexing}
\newacronym{tdd}{TDD}{time-division duplexing}
\newacronym{fdma}{FDMA}{frequency-division multiple access}
\newacronym{tdma}{TDMA}{time-division multiple access}
\newacronym{sdma}{SDMA}{space-division multiple access}
\newacronym[plural=MPCs]{mpc}{MPC}{multipath component}
\newacronym{mui}{MUI}{multi-user interference}

\newacronym{qam}{QAM}{quadrature amplitude modulation}
\newacronym{mqam}{MQAM}{M-ary quadrature amplitude modulation}

\newacronym{ls}{LS}{least-squares}
\newacronym{lms}{LMS}{least mean squares}
\newacronym{rls}{RLS}{recursive least-squares}
\newacronym{rzf}{RZF}{regularized zero-forcing}
\newacronym{mmse}{MMSE}{minimum mean square error}
\newacronym{lmmse}{LMMSE}{linear minimum mean square error}
\newacronym{mse}{MSE}{mean square error}
\newacronym{fft}{FFT}{fast Fourier transform}
\newacronym{dft}{DFT}{discrete Fourier transform}
\newacronym{dtft}{DTFT}{discrete-time Fourier transform}
\newacronym{ctft}{CTFT}{continuous-time Fourier transform}
\newacronym{ml}{ML}{machine learning}
\newacronym[plural=NNs]{nn}{NN}{neural network}
\newacronym[plural=RNNs]{rnn}{RNN}{recurrent neural network}
\newacronym[plural=ADCs]{adc}{ADC}{analog-to-digital converter}
\newacronym[plural=DACs]{dac}{DAC}{digital-to-analog converter}
\newacronym[plural=FPGAs]{fpga}{FPGA}{field-programmable gate array}
\newacronym{evm}{EVM}{error vector magnitude}
\newacronym{enob}{ENOB}{effective number of bits}
\newacronym{zf}{ZF}{zero-forcing}
\newacronym{rv}{r.v.}{random variable}
\newacronym{omp}{OMP}{orthogonal matching pursuit}
\newacronym{svd}{SVD}{singular value decomposition}

\newacronym{sdp}{SDP}{semidefinite programming}
\newacronym{psd}{PSD}{positive semidefinite}
\newacronym{nsd}{NSD}{negative semidefinite}

\newacronym{agc}{AGC}{automatic gain control}
\newacronym{rf}{RF}{radio frequency}
\newacronym{los}{LOS}{line-of-sight}
\newacronym{nlos}{NLOS}{non-line-of-sight}
\newacronym{ple}{PLE}{path loss exponent}
\newacronym[plural=dB,firstplural=decibels (dB)]{db}{dB}{decibel}
\newacronym[plural=dBm,firstplural=decibel milliwatts (dBm)]{dbm}{dBm}{decibel milliwatts}
\newacronym{pa}{PA}{power amplifier}
\newacronym{lna}{LNA}{low noise amplifier}
\newacronym{cw}{CW}{continuous wave}
\newacronym{papr}{PAPR}{peak-to-average power ratio}
\newacronym{usrp}{USRP}{Universal Software Radio Peripheral}
\newacronym{irr}{IRR}{image rejection ratio}
\newacronym{lo}{LO}{local oscillator}
\newacronym{vm}{VM}{vector modulator}
\newacronym{mmwave}{mmWave}{millimeter wave}
\newacronym{eirp}{EIRP}{effective isotropic radiated power}

\newacronym{csma}{CSMA}{carrier-sense multiple access}
\newacronym{csmaca}{CSMA/CA}{carrier-sense multiple access with collision avoidance}
\newacronym{csmacd}{CSMA/CD}{carrier-sense multiple access with collision detection}
\newacronym{mac}{MAC}{medium access control}
\newacronym{phy}{PHY}{physical layer}
\newacronym{4g}{4G}{fourth generation}
\newacronym{lte}{LTE}{Long-Term Evolution}
\newacronym{4glte}{4G LTE}{\gls{4g} \gls{lte}}
\newacronym{5g}{5G}{fifth generation}
\newacronym{nr}{NR}{New Radio}
\newacronym{5gnr}{5G NR}{5G New Radio}
\newacronym{ieee}{IEEE}{Institute of Electrical and Electronics Engineers}
\newacronym{wifi}{Wi-Fi}{IEEE 802.11}
\newacronym{lan}{LAN}{local area network}
\newacronym{wlan}{WLAN}{wireless local area network}
\newacronym[plural=BSs]{bs}{BS}{base station}
\newacronym[plural=SBSs]{sbs}{SBS}{small-cell base station}
\newacronym[plural=FD-SBSs]{fdsbs}{FD-SBS}{\gls{fd}-enabled \gls{sbs}}
\newacronym[plural=MBSs]{mbs}{MBS}{macrocell base station}
\newacronym[plural=UEs]{ue}{UE}{user equipment}
\newacronym{ul}{UL}{uplink}
\newacronym{dl}{DL}{downlink}
\newacronym{qos}{QoS}{Quality of Service}
\newacronym{fcc}{FCC}{Federal Communications Commission}
\newacronym{iab}{IAB}{integrated access and backhaul}
\newacronym{fab}{FAB}{fixed access and backhaul}
\newacronym{hetnet}{HetNet}{heterogeneous network}

\newacronym{siso}{SISO}{single-input single-output}
\newacronym{mimo}{MIMO}{multiple-input multiple-output}
\newacronym{sumimo}{SU-MIMO}{single-user \gls{mimo}}
\newacronym{mumimo}{MU-MIMO}{multi-user \gls{mimo}}
\newacronym{bf}{BF}{beamforming}
\newacronym{ca}{CA}{constant amplitude}
\newacronym{ula}{ULA}{uniform linear array}
\newacronym[\glslongpluralkey={angles of arrival}]{aoa}{AoA}{angle of arrival}
\newacronym[\glslongpluralkey={angles of departure}]{aod}{AoD}{angle of departure}
\newacronym{dof}{DoF}{degrees of freedom}
\newacronym{csi}{CSI}{channel state information}
\newacronym{csit}{CSIT}{\gls{csi} at the transmitter}
\newacronym{csir}{CSIR}{\gls{csi} at the receiver}
\newacronym{cs}{CS}{compressed sensing}

\newacronym{fd}{FD}{in-band full-duplex}
\newacronym{hd}{HD}{half-duplex}
\newacronym{si}{SI}{self-interference}
\newacronym{sic}{SIC}{self-interference cancellation}
\newacronym{soi}{SoI}{signal of interest}
\newacronym{asic}{A-SIC}{analog \acrlong{sic}}
\newacronym{dsic}{D-SIC}{digital \gls{sic}}
\newacronym{star}{STAR}{simultaneous transmit and receive}
\newacronym{warp}{WARP}{Wireless Open-Access Research Platform}
\newacronym{bfc}{BFC}{beamforming cancellation}
\newacronym{ipi}{IPI}{inter-panel-interference}
\newacronym{ipic}{IPIC}{inter-panel-interference cancellation}

\newacronym{qcqp}{QCQP}{quadratically-constrained quadratic programming}
\newacronym{cdf}{CDF}{cumulative density function}

\newacronym{elf}{ELF}{extremely low frequency}
\newacronym{slf}{SLF}{super low frequency}
\newacronym{ulf}{ULF}{ultra low frequency}
\newacronym{vlf}{VLF}{very low frequency}
\newacronym{lf}{LF}{low frequency}
\newacronym{mf}{MF}{medium frequency}
\newacronym{hf}{HF}{high frequency}
\newacronym{vhf}{VHF}{very high frequency}
\newacronym{uhf}{UHF}{ultra high frequency}
\newacronym{shf}{SHF}{super high frequency}
\newacronym{ehf}{EHF}{extremely high frequency}
\newacronym{thf}{THF}{tremendously high frequency}

\newacronym{wncg}{WNCG}{Wireless Networking and Communications Group}
\newacronym{linc}{LINC}{Laboratory of Informatics, Networks, and Communications}
\newacronym{ut}{UT Austin}{The University of Texas at Austin}
\newacronym{uiuc}{UIUC}{University of Illinois at Urbana-Champaign}
\newacronym{usc}{USC}{University of Southern California}
\newacronym{mit}{MIT}{Massachusetts Institute of Technology}
\newacronym{berkeley}{UC Berkeley}{University of California, Berkeley}
\newacronym{osu}{OSU}{Ohio State University}

\newcommand{\mmwave}{\gls{mmwave}\xspace}
\newcommand{\mimo}{\gls{mimo}\xspace}

\newcommand{\rf}{\gls{rf}\xspace}

\newcommand{\lmmse}{\gls{lmmse}\xspace}

\newcommand{\lna}{\gls{lna}\xspace}
\newcommand{\pa}{\gls{pa}\xspace}
\newcommand{\lnas}{\glspl{lna}\xspace}
\newcommand{\pas}{\glspl{pa}\xspace}

\newcommand{\adc}{\gls{adc}\xspace}
\newcommand{\adcs}{\glspl{adc}\xspace}

\newcommand{\secref}[1]{Section~\ref{#1}}

\newcommand{\figref}[1]{\figurename~\ref{#1}}
\newcommand{\algref}[1]{Algorithm~\ref{#1}}

\newcommand{\thmref}[1]{Theorem~\ref{#1}}

\begin{document}
	
%
\title{Hybrid Beamforming for Millimeter Wave Full-Duplex under Limited Receive\\Dynamic Range}
%
%
%

\author{%
	Ian~P.~Roberts,	Jeffrey~G.~Andrews,	and Sriram Vishwanath
	\thanks{I.~P.~Roberts and J.~G.~Andrews are with the Wireless Networking and Communications Group at the University of Texas at Austin, Austin, Texas, USA. S.~Vishwanath is with GenXComm, Inc., Austin, Texas, USA. Last revised: \today.}%
}

\maketitle




\begin{abstract}
Full-duplex \mmwave communication has shown increasing promise for self-interference cancellation via hybrid precoding and combining.
This paper proposes a novel \mmwave \mimo design for configuring the analog and digital beamformers of a full-duplex transceiver. 
Our design is the first to holistically consider the key practical constraints of analog beamforming codebooks, a minimal number of \rf chains, limited channel knowledge, beam alignment, and a limited receive dynamic range.
To prevent self-interference from saturating the receiver of a full-duplex device having limited dynamic range, our design addresses saturation on a per-antenna and per-\rf chain basis. 
Numerical results evaluate our design in a variety of settings and validate the need to prevent receiver-side saturation.
These results and the corresponding insights serve as useful design references for practical full-duplex \mmwave transceivers. 
\end{abstract}




\glsresetall


\section{Introduction} \label{sec:introduction}

The ability for a transceiver to transmit and receive simultaneously in-band introduces an exciting upgrade at the physical layer and in medium access when compared to existing half-duplex schemes such as \gls{tdd} and \gls{fdd} \cite{sabharwal_-band_2014}.
The gains supplied by full-duplex capability in \mmwave systems are particularly attractive \cite{xia_2017,roberts_wcm}, beyond the usual gains in spectral efficiency and latency.
By full-duplexing access and backhaul, heterogeneous \mmwave networks can be deployed with lower latency, higher spectral efficiency, and a reduced number of fiber drops.
Key challenges in \mmwave systems, such as beam alignment and beam tracking, have the potential to be transformed when devices can transmit and receive simultaneously, especially in highly dynamic environments like vehicle-to-vehicle communication.  
The presence of communication, radar, and other incumbents in lightly regulated \mmwave spectrum highlights the potential of novel strategies for medium access, in-band coexistence, and interference management via full-duplex. 

\subsection{Prior Work and Motivation}

The majority of existing research on full-duplex has been in the context of lower carrier frequencies, for example sub-6 GHz.
While many aspects of this existing work can be extended to \mmwave, new approaches are necessary to enable full-duplex at \mmwave \cite{xia_2017,roberts_wcm}.
The dense antenna arrays and wide bandwidths are two key challenges for analog self-interference cancellation at \mmwave in particular \cite{roberts_wcm}. 
Furthermore, directly translating \mimo-based self-interference mitigation (e.g., \cite{Huberman_Le-Ngoc_2015,everett_softnull_2016,Alexandropoulos_Duarte_2017}) to \mmwave is complicated by hybrid digital/analog beamforming, propagation characteristics at \mmwave, and system-level factors like beam alignment.
While passive and polarization-based approaches have been proposed for \mmwave \cite{dinc_60_2016,dinc_2017,singh_2020_acm}, they are difficult to generalize to dense \mmwave antenna arrays.

A number of recent works have investigated methods where self-interference is mitigated by appropriately configuring the transmit and receive beamformers at a \mmwave full-duplex device, sometimes termed \emph{beamforming cancellation} \cite{satyanarayana_hybrid_2019,liu_beamforming_2016,lopez_prelcic_2019_analog,lopez-valcarce_beamformer_2019,prelcic_2019_hybrid,roberts_2019_bfc,cai_robust_2019,roberts_2020_fsbfc,roberts_equipping_2020,zhu_uav_joint_2020}.
Resembling \mimo-based approaches from the sub-6 GHz full-duplex literature, existing beamforming cancellation designs  suggest that \mmwave full-duplex is theoretically possible without the hardware and computational costs associated with analog and digital self-interference cancellation.
Existing beamforming designs for \mmwave full-duplex, however, often fail to account for critical practical transceiver-level and system-level considerations.
To start, practical systems typically rely on codebook-based analog beamforming and beam alignment \cite{heath_overview_2016}, meaning there is extremely limited freedom in choosing analog beamformers. 
Designs such as those in \cite{liu_beamforming_2016,satyanarayana_hybrid_2019,prelcic_2019_hybrid,lopez_prelcic_2019_analog,lopez-valcarce_beamformer_2019,zhu_uav_joint_2020} do not account for codebook-based analog beamforming and assume the ability to fine-tune each phase shifter in analog beamforming networks.
Moreover, designs in \cite{liu_beamforming_2016,satyanarayana_hybrid_2019,lopez_prelcic_2019_analog,prelcic_2019_hybrid,lopez-valcarce_beamformer_2019,zhu_uav_joint_2020} assume infinite-precision phase shifters; in reality, phase shifters are almost certainly configured digitally, subjecting them to some degree of phase resolution.
Understandably, almost all designs assume a lack of amplitude control in analog beamforming even though it is not uncommon to have both phase and amplitude control in practice.
Those in \cite{liu_beamforming_2016,satyanarayana_hybrid_2019,lopez_prelcic_2019_analog,roberts_2019_bfc,prelcic_2019_hybrid,lopez-valcarce_beamformer_2019,zhu_uav_joint_2020} do not account for beam alignment and assume full over-the-air channel knowledge.

Several designs \cite{liu_beamforming_2016,lopez_prelcic_2019_analog,lopez-valcarce_beamformer_2019,zhu_uav_joint_2020} involve analog-only beamforming, meaning they only support single-stream communication, which simplifies the design of beamforming-based self-interference mitigation.
This is especially true in \cite{liu_beamforming_2016,satyanarayana_hybrid_2019,lopez_prelcic_2019_analog,prelcic_2019_hybrid,lopez-valcarce_beamformer_2019,zhu_uav_joint_2020} where the designs may be highly dependent on near-field self-interference channel conditions and are not shown to be robust against such.
Some designs, such as those in \cite{prelcic_2019_hybrid,roberts_2019_bfc,roberts_2020_fsbfc,roberts_equipping_2020}, take advantage of an increased number of \rf chains that allows them to exploit the consequent dimensionality to mitigate self-interference in the digital domain.
This is a strong assumption since the minimal number of \rf chains necessary in hybrid beamforming is equal to the number of streams; increasing beyond this is undesirable in terms of financial cost, size, and power consumption.

Finally, and perhaps most pertinent to this work, the majority of existing designs neglect the limited dynamic range of practical receivers \cite{rf_gu_2005}.
This is particularly important for full-duplex transceivers since self-interference---which is likely many orders of magnitude stronger than a desired receive signal---can saturate a receive chain if not sufficiently mitigated \cite{Day_Margetts_Bliss_Schniter_2012}.
The work in \cite{liu_beamforming_2016,lopez_prelcic_2019_analog,prelcic_2019_hybrid,lopez-valcarce_beamformer_2019,zhu_uav_joint_2020} accounts for \adc saturation by \textit{completely} mitigating self-interference beforehand but do not account for other sources of saturation (e.g., \lnas) and is not always possible. 
In \cite{roberts_equipping_2020}, the need to prevent \adc saturation is discussed, but this is assumed to be satisfied without any mathematical basis.
In \cite{satyanarayana_hybrid_2019,roberts_2019_bfc,roberts_2020_fsbfc}, the need to prevent receiver-side saturation is ignored.

\comment{
---

With all of this in mind, there exist several motivations for this work, in which we present a hybrid beamforming design for \mmwave full-duplex.
First and foremost, our design addresses receiver-side saturation at two stages in the hybrid beamforming architecture: (i) at each antenna and (ii) at each \rf chain.
To craft our design, we assume limited over-the-air channel knowledge and support codebook-based analog beamforming and beam alignment.
Furthermore, we restrict our deisng 

---

Most, if not all, other designs that we are aware of neglect the limited dynamic range of practical receivers altogether.
our design accounts for the limited dynamic range of a practical full-duplex receiver.
Most existing designs neglect this altogether or use one or more of the aforementioned impractical assumptions to circumvent it.
We are not aware of any existing work on \mmwave full-duplex that addresses receive dynamic range to the extent we have.

---

A significant differentiator of our design versus existing ones is that it does not demand more \rf chains than necessary.
While other designs may employ twice as many \rf chains as necessary or more, we do not, which is certainly more favorable in terms of size, financial cost, and power consumption.

---

In addition, some designs (e.g., CITE) exploit the structure or characteristics of a particular self-interference channel model; while there exist sensible, theory-based models for the self-interference channel at \mmwave, measurements have yet to confirm that a particular model holds in practice, making reliance on a particular model currently impractical.

---

This motivates the design presented herein, which aims to address key practical considerations while taking an information-theoretic approach.
By doing so, our design can be used in practical \mmwave full-duplex systems or, at the very least, as inspiration for future---perhaps even more practically sound---designs.
Furthermore, its numerical results can be to outline more realistic bounds on the capabilities of beamforming-based self-interference mitigation in practical \mmwave full-duplex systems.

Accounting for these artifacts of analog beamforming, our design relies solely on codebook-based analog beamforming and supports beam alignment, rather than assuming fine-tuned control of analog beamforming weights and knowledge of the over-the-air channel between our full-duplex transceiver and the devices it is serving; our design's formulation is agnostic to the particular codebook being used meaning it supports quantized phase and amplitude control.


Thus, we were motivated to construct a design that does not rely on a particular self-interference channel model.

Finally, and perhaps most importantly, our design accounts for the limited dynamic range of a practical full-duplex receiver.
Most existing designs neglect this altogether or use one or more of the aforementioned impractical assumptions to circumvent it.
We are not aware of any existing work on \mmwave full-duplex that addresses receive dynamic range to the extent we have.
Our methodology ensures that the \mmwave \mimo design herein mitigates self-interference---on a per-antenna basis and a per-\rf chain basis---such that the receiver does not saturate, which could otherwise severely degrade reception of a desired receive signal.
In particular, we motivate our design by gain compression exhibited by practical \lnas, which are commonly per-antenna, and the limited resolution of \adcs, which are per-\rf chain.
}

\subsection{Contributions}

We formulate \mmwave \mimo expressions capturing practical receive dynamic range limitations per-antenna and per-\rf chain.
In particular, we motivate this work by the limited dynamic range of \lnas placed per-antenna and of \adcs placed per-\rf chain.
Using these formulations, we outline constraints on a \mmwave \mimo design to limit the self-interference power inflicted on each antenna and each \rf chain at the receiver of the full-duplex device.
By doing so, we can ensure a limited receive dynamic range does not severely degrade reception of a desired signal.
We outline conditions where meeting these per-antenna and per-\rf chain self-interference power constraints are implicitly met, either by one another or other system factors.

We present a hybrid digital/analog beamforming design that enables a \mmwave transceiver to operate in a full-duplex fashion, serving two devices simultaneously in-band.
Our design aims to achieve a high sum spectral efficiency on the two links while ensuring it does not induce receiver-side saturation on a per-antenna and per-\rf chain basis.
Adhering to a multitude of additional practical considerations beyond a limited receive dynamic range, our design supports beam alignment schemes and codebook-based analog beamforming, rather than assuming full knowledge of the over-the-air channels and the ability to fine-tune phase shifters and attenuators. Furthermore, we limit the number of \rf chains to the minimum necessary for multi-stream transmission. 
To provide our design with some freedom in the choice of its analog beamformers, we present a methodology for building sets of candidate analog beamformers based on measurements from codebook-based beam alignment.
Finally, our design is not self-interference channel model-dependent, in that it does not exploit any particular structure or model.

We evaluate our design under a variety of settings. 
Our numerical results indicate scenarios where our design thrives, offering significant spectral efficiency gains over conventional half-duplex operation. 
These results also outline conditions under which per-antenna and per-\rf chain self-interference power constraints restrict what is possible for \mmwave full-duplex, providing useful insights to engineers on relationships between system parameters such as transmit power, \rf isolation, \adc resolution, and the self-interference power reaching each antenna and each \rf chain. 
Understanding the degree of self-interference mitigation required at specific points in the receiver can drive full-duplex system analyses, including those that may supplement beamforming-based approaches with analog and/or digital self-interference cancellation \cite{roberts_equipping_2020}.

\comment{
---

The main contributions of this work are as listed as follows.
\begin{itemize}
	\item We provide formulations and calculations linking single-letter \mmwave \mimo expressions to practical per-antenna and per-\rf chain dynamic range considerations. In particular, we motivate our design by the limited dynamic range of \lnas placed per-antenna and of \adcs placed per-\rf chain.
	\item We formulate the problem of designing a hybrid digital/analog beamforming configuration that enables a full-duplex transceiver to serve two half-duplex devices simultaneously in-band while preventing self-interference from inducing receiver-side saturation on a per-antenna and per-\rf chain basis. Our approach to solving this problem is driven by information-theoretic motivations and hard constraints on the amount of self-interference inflicted onto the receiver of the full-duplex device.
	\item Adhering to a multitude of additional practical considerations, our design supports beam alignment schemes and codebook-based analog beamforming, rather than assuming full knowledge of the over-the-air channels and the ability to fine-tune phase shifters and attenuators. Furthermore, we limit the number of \rf chains to the minimum necessary for multi-stream transmission (i.e., the number of \rf chains is equal to the number of streams). Finally, our design is not self-interference channel model-dependent, in that it does not exploit any structure or properties of such.
	\item We outline conditions where per-antenna and per-\rf chain constraints are implicitly met, which can be used to simplify and accelerate our design and provide insights on the relationship between the constraints.
	\item To provide our design with some freedom in the choice of its analog beamformers, we present a methodology for building sets of candidate analog beamformers based on a flexible number of measurements from codebook-based beam alignment. 
	\item We evaluate our design under a variety of settings. These numerical results indicate scenarios where our design thrives, offering significant spectral efficiency gains over conventional half-duplex operation. Our numerical results also outline conditions under which per-antenna and per-\rf chain constraints restrict what is possible for \mmwave full-duplex, providing useful insights to engineers on relationships between system parameters such as transmit power, \rf isolation, \adc resolution, and the self-interference power reaching each antenna and each \rf chain. Understanding the degree of self-interference mitigation required at specific points in the receiver can drive full-duplex system analyses, including those that may supplement beamforming-based approaches with analog and/or digital self-interference cancellation \cite{roberts_equipping_2020}.
\end{itemize}
}

\section{System Model} \label{sec:system-model}

\begin{figure*}[!t]
	\centering
	\includegraphics[width=\linewidth,height=0.35\textheight,keepaspectratio]{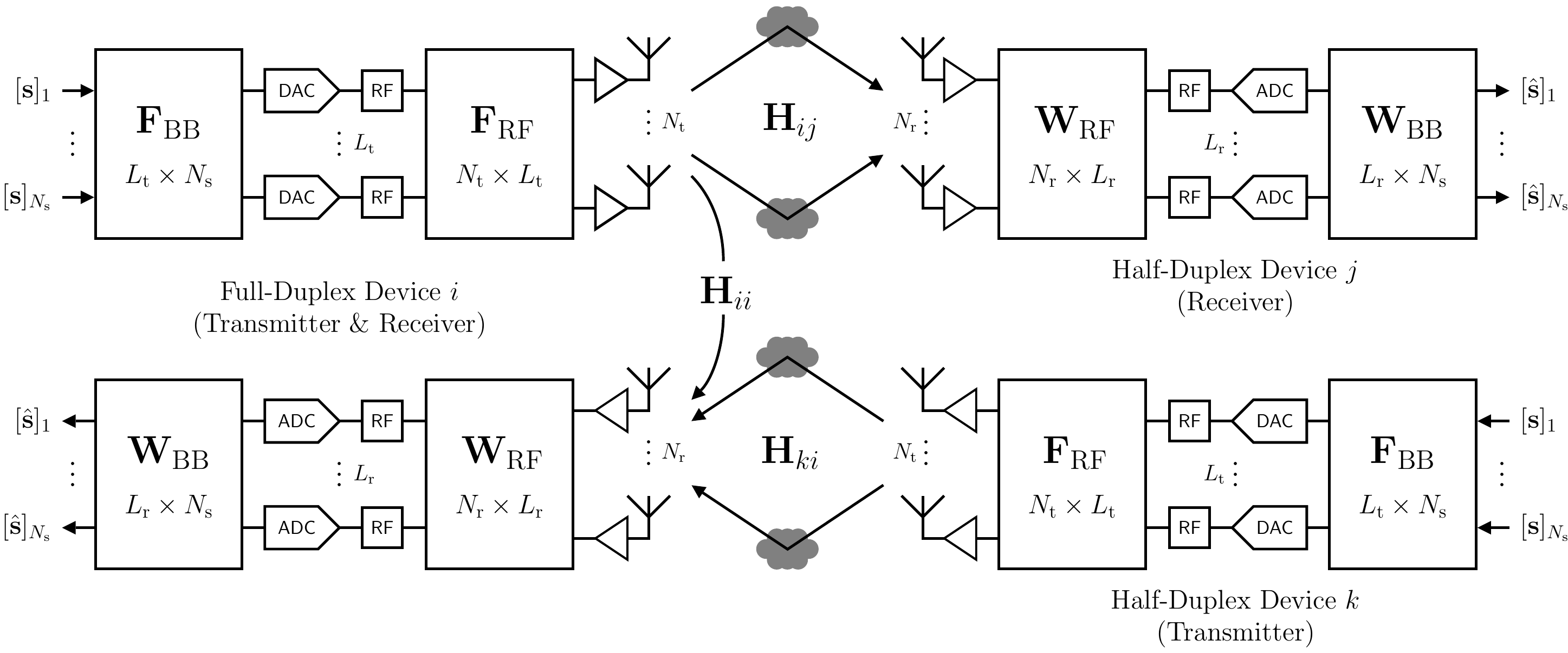}
	\caption{A full-duplex \mmwave device $i$ transmitting to $j$ as it receives from $k$ in-band.} 
	\label{fig:network}
\end{figure*}


This work considers the wireless system in \figref{fig:network}, where a \mmwave transceiver $i$ aims to transmit to a device $j$ while receiving from a device $k$ in the same band.
Instead of turning to half-duplexing strategies like \gls{tdd} or \gls{fdd} to avoid self-interference, this work presents a design that enables in-band full-duplex operation by leveraging the spatial domain to mitigate self-interference.
It is important to note that devices $j$ and $k$ could be separate half-duplex devices, separate full-duplex devices, or a single full-duplex device.
In this work, we consider the general case where they are separate half-duplex devices, though many aspects of our contribution would extend naturally, or even simplify, under the other settings.
 

Ubiquitous among practical \mmwave transceivers to efficiently harness dense antenna arrays is the use of hybrid digital/analog beamforming architectures where transmit precoding and receive combining are implemented by the combination of digital (baseband) and analog (\rf) signal processing, as exhibited in \figref{fig:network} \cite{heath_overview_2016}.
We assume devices $i$, $j$, and $k$ all employ hybrid beamforming in a fully-connected fashion where each antenna is connected to each \rf chain via an analog beamforming network. 
As illustrated in \figref{fig:network}, we assume that separate arrays are used at device $i$ for transmission and reception and independent precoding and combining on the two is supported.
This assumption is motivated by the fact that \mmwave circulators that are sufficient for full-duplex operation are still out of reach \cite{dinc_2017}.

For $(m,n) \in \{(i,j),(k,i)\}$, we use the following notation. 
Let $\Nt\node{m}$ and $\Nr\node{n}$ be the number of transmit and receive antennas, respectively. 
Connecting the digital and analog stages, let $\Lt\node{m}$ and $\Lr\node{n}$ be the number of transmit and receive \rf chains, respectively. 
Let $\Ns\node{mn}$ be the number of symbol streams transmitted from device $m$ intended for device $n$. 
Let $\prebb\node{m} \in \complexpow{\Lt\node{m}}{\Ns\node{mn}}$ be the digital precoding matrix and $\prerf\node{m} \in \complexpow{\Nt\node{m}}{\Lt\node{m}}$ be the analog precoding matrix, responsible for transmitting from $m$. 
Let $\combb\node{n} \in \complexpow{\Lr\node{n}}{\Ns\node{mn}}$ be the digital combining matrix and $\comrf\node{n} \in \complexpow{\Nr\node{n}}{\Lr\node{n}}$ be the analog combining matrix, responsible for receiving at $n$. 


%
%

For $(m,n) \in \{(i,j),(k,i)\}$, let $\symbvec\node{m} \in \setvectorcomplex{\Ns\node{mn}}$ be the symbol vector transmitted by $m$ intended for device $n$, where the symbol covariance is
\begin{align}
\ev{\symbvec\node{m}\symbvec\ctransnode{m}} = \frac{1}{\Ns\node{mn}} \ \mat{I} \label{eq:symbol-covariance}.
\end{align}
We do not consider a specific signaling constellation, though we will evaluate our work assuming Gaussian signaling is employed.
Let $\noisevec\node{n} \sim \cgauss{\mat{0}}{\noisevar \cdot \mat{I}}$ be the $\Nr\node{n} \times 1$ additive noise vector incurred at the receive array of $n$, where $\noisevar$ represents a per-antenna noise power in watts/Hz and is assumed common across devices for simplicity.

We denote the symbol period as $\symboltime$ and the symbol bandwidth $B = \symboltime\inv$,
which we assume to be constant across both links, being a full-duplex system.
Let $\Ptxwatts\node{m}$ be the total transmit power of device $m$ in joules per second (watts).
Let $\Ptx\node{m}$ be the resulting transmit power in joules per symbol.
We extend this convention, representing power quantities in watts using a tilde, $\tilde{P}$, and in joules per symbol as without a tilde, $P$, which are linked via $P = \tilde{P} \cdot B\inv = \tilde{P} \cdot \symboltime$.

We impose the following digital precoding power constraint 
\begin{gather}
\normfro{\prebb\node{m}}^2 \leq 1 \label{eq:baseband-precoder-power-constraint}
\end{gather}
and normalize the columns of $\prerf\node{m}$ to have squared $\ell_2$-norm $\Lt\node{m}$ and of $\comrf\node{n}$ to have squared $\ell_2$-norm $\Nr\node{n}$.
Since this work is focused on receiver-side power levels, our analog combining constraint differs from the analog precoding power constraint to ensure consistency with the physical combining taking place at receivers.
As is common in practice, we will assume that columns of our analog precoders and combiners will come from analog beamforming codebooks that account for hardware constraints such as phase shifter resolution and amplitude control.
That is, for $(m,n) \in \{(i,j),(k,i)\}$, we have
\begin{gather}
\entry{\prerf\node{m}}{:,\ell} \in \prerfcb\node{m}, \ \ell = 1, \dots, \Lt\node{m} \label{eq:rf-codebook-i-pre} \\
\entry{\comrf\node{n}}{:,\ell} \in \comrfcb\node{n}, \ \ell = 1, \dots, \Lr\node{n} \label{eq:rf-codebook-i-com} 
\end{gather}
where $\prerfcb\node{m}$ and $\comrfcb\node{n}$ denote analog precoding and combining codebooks, respectively. 
Now, let us consider $(m,n) \in \{(i,j),(k,i),(i,i)\}$.
We assume that the large-scale power gain between devices $m$ and $n$ is given by $G^2_{mn}$.
The $\Nr\node{n} \times \Nt\node{m}$ channel matrix between a transmitter $m$ and receiver $n$ is denoted $\channel_{mn} \in \complexpow{\Nr\node{n}}{\Nt\node{m}}$.
In this work, we consider the more straightforward case of frequency-flat \mimo channels and will address frequency-selective ones in future work.
Taking the perspective of our full-duplex device $i$, we term $\channel_{ij}$ the \textit{transmit} channel, $\channel_{ki}$ the \textit{receive} channel, and $\channel_{ii}$ the \textit{self-interference} channel.
Note that we have not considered an inter-user interference channel between devices $k$ and $j$ since we assume that, with sufficient separation, the interference between the two to be negligible given the high path loss at \mmwave and highly directional steering of energy that is typical.

We assume devices $i$ and $j$ as well as devices $k$ and $i$ are separated in a far-field fashion. 
As such, we model the transmit and receive channels with the Saleh-Valenzuela-based representation where propagation from one device to another is modeled by the composition of several discrete rays. 
Explicitly, channels $\channel_{ij}$ and $\channel_{ki}$ are modeled as follows \cite{heath_overview_2016}, where $(m,n) \in \{(i,j),(k,i)\}$, 
\begin{align}
\channel_{mn} = 
\sqrt{\frac{1}{\numrays\node{mn}}}
\sum_{u=1}^{\numrays\node{mn}} 
\beta_{u} \
\rxresponsevector\node{n}(\aoa_{u}) \ 
\txresponsevector\node{m}(\aod_{u})\ctrans. \label{eq:desired-channel}
\end{align}
In each channel, $\numrays\node{mn}$ is a random variable dictating the number of rays in the channel. 
The complex gain of ray $u$ is given as $\beta_{u} \sim \cgauss{0}{1}$. 
The $u$-th ray's \gls{aod} and \gls{aoa} are given as $\aod_{u}$ and $\aoa_{u}$, respectively.
The transmit and receive array response vectors at these angles are given as $\txresponsevector\node{m}(\aod_{u})$ and $\rxresponsevector\node{n}(\aoa_{u})$, which have squared $\ell_2$-norm $\Nt\node{m}$ and $\Nr\node{n}$, respectively.
The coefficient in front of the summations handles a channel power normalization to ensure $\ev{\frobeniustwo{\channel_{mn}}} = \Nt\node{m}\Nr\node{n}$.

The self-interference channel $\channel_{ii}$ presents its own unique modelling questions \cite{roberts_wcm}.
A lack of measurements and characterization of such a channel prevents us from confidently assuming a particular channel model.
As such, our contribution herein does not rely on the self-interference channel's structure or properties.
However, to evaluate our design, we employ a model that aims to capture the near-field nature of the transmit and receive arrays at $i$ along with reflections that may stem from the environment \cite{xia_2017,satyanarayana_hybrid_2019}.
We explicitly state this model in \secref{sec:simulation-results}.
The large-scale power gain of the self-interference channel is represented by $G^2_{ii}$, which captures the \rf isolation between the transmit and receive arrays at $i$.
We define the following \gls{snr} between two devices $(m,n) \in \{(i,j),(k,i)\}$ as
\begin{align}
\snr_{mn} 
&\triangleq \frac{\Pt\node{m} G^2_{mn}}{\noisevar} \label{eq:snr-b} = \frac{\Ptxwatts\node{m} G^2_{mn}}{\noisevar \cdot B}
\end{align}
which captures the received power (without beamforming gains) versus the noise power. 

\comment{
\subsection{Receiver-Side Components}
In this work, given that we will address \pa and \lna saturation, the placement of these components is important to declare before presenting our design.
As indicated by \figref{fig:transmitter} and \figref{fig:receiver}, we consider the particular case where a transmitter's \pas are per-antenna and a receiver's \lnas are per-antenna. 
Moreover, a receiver's \adcs are per-\rf chain.

This work seeks to address \lna and \adc saturation at the receiver of the full-duplex device $i$.
When receiving while transmitting in-band, the self-interference introduced at $i$ combines with the desired receive signal from $k$.
Practically, it is safe to assume the self-interference is much stronger than a desired receive signal: a desired signal is likely to have propagated tens or hundreds of meters while the self-interference has traveled mere centimeters.
The desired signal and self-interference combine at the receive antennas, introducing issues twofold:
\begin{enumerate}
	\item the self-interference can push operation of the \lnas beyond their linear region
	\item the self-interference can saturate the \adcs, reducing the effective number of bits offered to quantization of the desired signal
\end{enumerate}

A calculation of \etalna is presented in Section~\ref{sec:deriving-lna-adc-constraints} and Section~\ref{sec:link-analysis}.
}





\section{Problem Formulation} \label{sec:problem-formulation}

This work is motivated by the fact that a receive chain of a full-duplex device---which practically has a limited dynamic range---is susceptible to saturation due to the overwhelming strength of self-interference \cite{rf_gu_2005}.
To highlight this, we consider two sources of limited dynamic range in this work: \lnas and \adcs.
Like other amplifiers, \lnas begin saturating and introduce significant nonlinearities beyond a certain input power level, meaning only signals below some power threshold see an approximately linear amplifier.
The limited resolution of an \adc is a classical example of limited dynamic range.
Since the combination of a desired receive signal and self-interference enters the \adc, self-interference can drive up quantization noise and degrade the quality of the desired receive signal.

Consider \figref{fig:network}, where \lnas are placed per-antenna and \adcs are placed per-\rf chain.
To avoid saturating the \lnas at the full-duplex device, the self-interference power reaching each antenna must be mitigated to below some threshold.
Similarly, to avoid saturating the \adcs, the self-interference reaching each \rf chain must also be limited.
This work investigates relying solely on beamforming to achieve full-duplex, meaning the design herein must prevent \lna and \adc saturation.
Note that we are only concerned with preventing saturation at the receiver of the \textit{full-duplex} device $i$---not that of the \textit{half-duplex} device $j$---since the saturation we are considering stems from self-interference.
Also, note that \lnas are not always placed strictly per-antenna as we have assumed and that there may exist other motivations for restricting the self-interference power at each antenna and at each \rf chain.
With this in mind, the design we present is not strictly for \lnas and \adcs but rather for meeting arbitrary per-antenna and per-\rf chain self-interference power constraints: \lnas and \adcs are an important special case.
Using the previously defined system model, we can begin analyzing the signals that reach the \lnas and \adcs of our full-duplex device.
The symbol vector at the \lnas of $i$ is
\begin{align}
\ylna = \ylnades + \ylnaint + \ylnanoise \in \setvectorcomplex{\Nr\node{i}}
\end{align}
where the desired term is $\ylnades = \sqrt{\Ptx\node{k}} G_{ki} \channel_{ki} \prerf\node{k} \prebb\node{k} \symbvec\node{k}$, the self-interference term is
$\ylnaint = \sqrt{\Ptx\node{i}} G_{ii} \channel_{ii} \prerf\node{i} \prebb\node{i} \symbvec\node{i}$, and the noise term is $\ylnanoise = \noisevec\node{i}$.
Following reception, the signal at each antenna passes through its respective \lna.
We represent \lna operation as $\lnafun{\cdot}$ and model it in the following piece-wise fashion
\begin{align} \label{eq:lna-piecewise}
\lnafun{x} = 
\begin{cases}
\lnafunlin{x} = \lnagain \cdot x, & \bars{x}^2 \leq \powertotmaxlna \\
\lnafunsat{x}, & \bars{x}^2 > \powertotmaxlna
\end{cases}
\end{align}
where the per-antenna symbol $x$ undergoes linear amplification $\lnafunlin{\cdot}$ offering a linear amplitude gain of $\lnagain > 0$ if its average power (over the symbol period) is below some threshold $\powertotmaxlna$.
Otherwise, the \lna is saturated and $x$ undergoes some nonlinear function $\lnafunsat{\cdot}$.
It is difficult to characterize $\lnafunsat{\cdot}$ at the symbol level since \lna saturation takes place instantaneously on time-domain signals.
For this reason, we make no attempt to characterize $\lnafunsat{\cdot}$ with the understanding that linear \lna operation can be ensured by restricting the power of $x$ up to $\powertotmaxlna$.
We would like to point out that for a properly chosen $\powertotmaxlna$---which will likely include appropriate backoffs for the signal distribution and pulse shape---the \textit{time-domain signal} will undergo linear amplification and, thus, so will the symbols.
Since the gain of the \lna acts on signal-plus-noise, it can be abstracted out henceforth as $\lnagain = 1$.


For ease of notation, we overload the \lna transfer function $\lnafun{\cdot}$ function to support vector input by the simple element-wise extension $\entry{\lnafun{\vx}}{\ell} = \lnafun{x_\ell}$,
where $\vx = [x_1, x_2, \dots]\trans$.
Following per-antenna \lna amplification, the signals from each antenna are combined as
\begin{align}
\yadc &= \comrf\ctransnode{i} \times \lnafun{\ylna} \in \setvectorcomplex{\Lr\node{i}} \label{eq:lna-output-1}
\end{align}
where $\yadc$ is the vector of per-\rf chain symbols reaching the \adcs.
Under linear \lna operation, \eqref{eq:lna-output-1} can be written as
\begin{align}
\yadc 
&\stackrel{\substack{\mathrm{lin}}}{=} \comrf\ctransnode{i} \times \ylna
= \yadcdes + \yadcint + \yadcnoise
\end{align}
where $\yadcdot = \comrf\ctransnode{i} \times \ylnadot$.

We use $\adcfun{b}{\cdot}$ to represent a $b$-bit \adc, which can be modelled as
\begin{align}
\adcfun{b}{x} = x + e_{\mathrm{quant}}
\end{align}
where $x$ is the symbol reaching the \adc and $e_{\mathrm{quant}}$ is the error in perfectly observing $x$ due to quantization noise.
A useful approximation of quantization noise power under $\adcbits$-bit, uniform quantization is \cite{rf_gu_2005}
\begin{align}
\ev{e_{\mathrm{quant}}^2} = \frac{8 \cdot \bars{x}^2}{12 \cdot 2^{2\adcbits}} \label{eq:power-quant-1}.
\end{align}
As the number of bits $b$ increases, the magnitude of $e_{\mathrm{quant}}$ decreases.
Similarly, as the magnitude of $x$ increases, the quantization noise power increases.
From this, one can see why mitigating self-interference before the \adc input is so important: increased self-interference can plague a desired receive signal with increased quantization noise.
We overload $\adcfun{b}{\cdot}$ to vectors as $\entry{\adcfun{b}{\vx}}{\ell} = \adcfun{b}{x_\ell}$, 
where $\vx = [x_1, x_2, \dots]\trans$.
The symbol vector out of the \adcs is
\begin{align}
\ydig 
&= \adcfun{b}{\yadc} 
= \yadcdes + \yadcint + \yadcnoise + \equant \in \setvectorcomplex{\Lr\node{i}}.
\end{align}
From this discussion, we can see that limiting the power into the \lnas and into the \adcs is critical in preserving the linearity of the receive chain and reducing the effects of quantization.
With these models and formulations in hand, we begin laying out our contribution, which aims to mitigate self-interference to certain levels per-antenna and per-\rf chain.









\section{Beam Candidate Acquisition} \label{sec:beams}


To avoid estimating the over-the-air channel and to address initial access, practical \mmwave systems employ \textit{beam training}, or \textit{beam alignment}, schemes that aim to identify transmit-receive beam pairs that afford link margin sufficient for communication.
Once promising beams have been identified and assigned, analog beamformers at the link's transmitter and receiver can be set, which nets an effective channel that can be more straightforwardly estimated and can sustain communication.
Typically, these analog beamformers are configured according to a predetermined codebook of beams, which reduces complexity and offers robustness.

In this section, we describe a \textit{beam candidate acquisition} algorithm used to construct a set of analog beamforming candidates when knowledge of the over-the-air channel (e.g., $\channel_{ij}$, $\channel_{ki}$) is not present.
Rather than restricting our hybrid beamforming design to transmit and receive along a single choice of analog beamformers, a set of analog beamforming candidates will be supplied to our design in \secref{sec:design}.
This is motivated by the fact that some transmit-receive beam pairs at the full-duplex device will naturally afford more isolation than other pairs, meaning it may be preferable to use them for full-duplex, even if they were sub-optimal in a half-duplex setting.




For our design, we propose that candidate analog beamforming matrices can be created as follows.
Let $\prerftr\node{i} \in \setmatrixcomplex{\Nt\node{i}}{\Mt\node{i}}$ be a matrix whose $\Mt\node{i}$ columns are training analog precoders used by $i$ during candidate beam acquisition as it illuminates $\channel_{ij}$. 
To observe these illuminations, let $\comrftr\node{j} \in \setmatrixcomplex{\Nr\node{j}}{\Mr\node{j}}$ be a matrix whose $\Mr\node{j}$ columns are training analog combiners used by $j$.
We assume, for simplicity, that each of the $\Mt\node{i}$ training precoders is observed by all $\Mr\node{j}$ training combiners, though the ideas herein could easily be extended when this is not the case.
Analogously, let $\prerftr\node{k} \in \setmatrixcomplex{\Nt\node{k}}{\Mt\node{k}}$ and $\comrftr\node{i} \in \setmatrixcomplex{\Nr\node{i}}{\Mr\node{i}}$ be the training analog precoders and analog combiners used by $k$ and $i$, respectively, used to measure $\channel_{ki}$.
On both links, we assume the analog beamformers used during training come from their respective codebooks according to \eqref{eq:rf-codebook-i-pre}--\eqref{eq:rf-codebook-i-com} and are highly directional and linearly independent. 
{With slight modifications, the ideas that follow could also accommodate cases where measurements map to codebook candidates rather than being the candidates themselves.}
The collection of measurements for each link can be written in matrix form as
\begin{align}
\mM_{ij} &= \sqrt{\Ptx\node{i}} G_{ij} \comrftr\ctransnode{j} \channel_{ij} \prerftr\node{i} \eyemat{\Mt\node{i}} \in \setmatrixcomplex{\Mr\node{j}}{\Mt\node{i}} \label{eq:beams-measured-ij} \\
\mM_{ki} &= \sqrt{\Ptx\node{k}} G_{ki} \comrftr\ctransnode{i} \channel_{ki} \prerftr\node{k} \eyemat{\Mt\node{k}} \in \setmatrixcomplex{\Mr\node{i}}{\Mt\node{k}} \label{eq:beams-measured-ki}.
\end{align}
Note that $\mM_{ij}$ may be measured in the reverse fashion, where $j$ transmits to $i$, to avoid feedback overhead. 
For consistency, we maintain notation as if measurements take place from $i$ to $j$.

Given that the training precoders and combiners come from their respective codebooks, the strength of the measurements in $\mM_{ij}$ and $\mM_{ki}$ directly indicates which analog precoders and analog combiners are promising candidates on each link.
When only one \rf chain is used (i.e., analog-only beamforming), the strongest entry in $\mM_{ij}$ suggests that it should be used for communication from $i$ to $j$ (and likewise on the link from $k$ to $i$).
Having multiple \rf chains allows us to transmit/receive with multiple beams at once to support spatial multiplexing.
We now look at how to build multiple analog beamforming candidates---that is, a set of pairs $\parens{\prerf\node{i},\comrf\node{j}}$ that support multi-stream communication on $\channel_{ij}$ and likewise from $k$ to $i$.


%

Let $\Tij$ be a set of $K_{ij}$ analog precoding-combining pairs shown in \eqref{eq:Tij} used for communication from $i$ to $j$.
Similarly, let $\Tki$ be a set of $K_{ki}$ analog precoding-combining pairs shown in \eqref{eq:Tki} used for communication from $k$ to $i$.
\begin{align}
\Tij &= \braces{\parens{\prerf\node{i},\comrf\node{j}} : \entry{\prerf\node{i}}{:,\ell_{\mathrm{t}}} \in \prerfcb\node{i} \ \forall \ \ell_{\mathrm{t}}, \entry{\comrf\node{j}}{:,\ell_{\mathrm{r}}} \in \comrfcb\node{j} \ \forall \ \ell_{\mathrm{r}}} \label{eq:Tij} \\
\Tki &= \braces{\parens{\prerf\node{k},\comrf\node{i}} : \entry{\prerf\node{k}}{:,\ell_{\mathrm{t}}} \in \prerfcb\node{k} \ \forall \ \ell_{\mathrm{t}}, \entry{\comrf\node{i}}{:,\ell_{\mathrm{r}}} \in \comrfcb\node{i} \ \forall \ \ell_{\mathrm{r}}} \label{eq:Tki}
\end{align}
Our goal is to form the candidate sets $\Tij$ and $\Tki$ with promising beamforming pairs for each link. 
We describe our method for doing so, which can be replicated for $\Tij$ and $\Tki$ independently, by using generic notation (e.g., $\set{T}$, $K$, $\mM$) according to the summary shown in \algref{alg:beams}.

We begin by finding the indices of the training precoders $\Mtx$ and of the training combiners $\Mrx$ that revealed the top $K$ strongest measurements in $\mM$, sorted according to descending strength.
We initialize $\mathcal{T}$ to an empty set.
The first beam (column) of the $k$-th candidate is steered along the $k$-th strongest entry in $\mM$. 
To ensure that we do not transmit or receive along directions more than once, we keep track of each beam's transmit and receive indices (i.e., $t$ and $r$) in sets $\Jtx$ and $\Jrx$, respectively.
Then, to choose the next column, we locate the strongest entry in $\mM$ whose transmit or receive beam has not already been selected for this $k$-th candidate.
When the columns of $\prerftr$ and $\comrftr$ are linearly independent, this ensures that each analog precoding candidate and analog combining candidate are rank-$L$, which is necessary for multiplexing up to $L$ streams.
Note that, in this method, we have assumed that $L = \Lt = \Lr = \Ns$ for a given link, which is more practical than supplying devices with more \rf chains than streams.
This process is repeated until all $L$ columns are populated, and then the candidate pair is appended to our candidate set $\mathcal{T}$.
Once all $K$ candidates have been generated, the set $\mathcal{T}$ is returned.

\begin{algorithm}[!t]
	\begin{algorithmic}[0]
		\REQUIRE $\mM$, $\prerftr$, $\comrftr$, $L$, $K$
		\STATE $\mathcal{T} = \braces{\varnothing}$
		\STATE $\Mrx,\Mtx = \arg\maxkop{\abs{\mM},K}$
		\FOR{$k = 1:K$}
			\STATE $t = \entry{\Mtx}{k}$, $r = \entry{\Mrx}{k}$
			\STATE $\prerf = \entry{\prerftr}{:,t}$
			\STATE $\comrf = \entry{\comrftr}{:,r}$
			\STATE $\mathcal{J}_{\mathrm{tx}} = \braces{t}$, $\mathcal{J}_{\mathrm{rx}} = \braces{r}$
			\FOR{$\ell = 1:L-1$}
				\STATE $t,r = \arg \max_{t,r} \bars{\entry{\mM}{t,r}} \ \st t \notin \mathcal{J}_{\mathrm{tx}}, \ r \notin \mathcal{J}_{\mathrm{rx}}$
				\STATE $\prerf = \brackets{\prerf \quad \entry{\prerftr}{:,t}}$
				\STATE $\comrf = \brackets{\comrf \quad \entry{\comrftr}{:,r}}$
				\STATE $\mathcal{J}_{\mathrm{tx}} = \mathcal{J}_{\mathrm{tx}} \cup t$, $\mathcal{J}_{\mathrm{rx}} = \mathcal{J}_{\mathrm{rx}} \cup r$
			\ENDFOR
			\STATE $\mathcal{T} = \mathcal{T} \cup \parens{\prerf,\comrf}$
		\ENDFOR
		\ENSURE $\mathcal{T}$
	\end{algorithmic}
	\caption{Beam candidate acquisition algorithm.}
	\label{alg:beams}
\end{algorithm}


Given the spatial sparsity of \mmwave channels, it is likely there will be discrete rays comprising the channel, as evidenced by \eqref{eq:desired-channel}.
Transmission and reception will take place along these rays, as there will be very little energy in other directions.
Therefore, to build promising candidates, it is critical that $\Mt$ and $\Mr$ be sufficiently large to locate at least $K$ strong rays in the channel.
We would like to point out that this proposed method, with appropriate modification, can easily be extended to beam training procedures that hierarchically locate rays or use compressed sensing to inspect the over-the-air channel.


Following the construction of $\Tij$ and $\Tki$ using this method, we now suggest that the following channel estimates be made.
We have not assumed knowledge of $\channel_{ij}$ or $\channel_{ki}$ (otherwise this beam candidate acquisition would be immaterial).
Therefore, to provide the design presented in the next section with channel information, we assume the following set $\mathcal{H}_{ij}$ has been populated with measurements of the effective channel seen by each of the candidates in $\Tij$ described as
\begin{align}
\mathcal{H}_{ij} = \braces{ \comrf\ctransnode{j} \channel_{ij} \prerf\node{i} : \parens{\prerf\node{i},\comrf\node{j}} \in \mathcal{T}_{ij} } \label{eq:rf-cand-channels-ij}
\end{align}
where $\card{\mathcal{H}_{ij}} = K_{ij}$.
Note that each effective channel in $\mathcal{H}_{ij}$ is merely $\Lr\node{j} \times \Lt\node{i}$, a very small size relative to $\channel_{ij}$, and is observed digitally (i.e., from the \glspl{dac} of $i$ to the \adcs of $j$).
By these two facts, it is our hope that the overhead associated with collecting the measurements in $\mathcal{H}_{ij}$ not be prohibitive.
Furthermore, we would like to point out that our design does not require executing this estimation on the link from $k$ to $i$.
This concludes beam candidate acquisition, having populated $\Tij$, $\Tki$, and $\mathcal{H}_{ij}$, which will enable the design presented in the next section.

\section{Hybrid Beamforming Design for mmWave Full-Duplex} \label{sec:design}

In this section, we present a hybrid beamforming design that will enable \mmwave full-duplex while accounting for per-antenna and per-\rf chain power constraints at the receiver of a full-duplex device.
The goal of our design is to achieve a high spectral efficiency on both links while ensuring that the power of the self-interference reaching the full-duplex device's receiver is below some thresholds (which we define shortly).
To do so, our design leverages the analog beamforming candidate sets $\Tij$ and $\Tki$ found in the previous section to configure the analog beamformers at each device.
We would like to remind the reader that the design that follows holds for general per-antenna and per-\rf chain power constraints, even though we are considering \lna and \adc power constraints as particular motivators.

Our design supports spatial multiplexing of multiple streams and importantly does not require more \rf chains than necessary, where $\Lt\node{i} = \Lr\node{j} = \Ns\node{ij}$ and $\Lt\node{k} = \Lr\node{i} = \Ns\node{ki}$.
Furthermore, we support the important case where $\Ns\node{ij} = \Ns\node{ki}$, implying $\Lt\node{i} = \Lr\node{i}$.
This, along with the fact that we have not assumed anything about $\channel_{ii}$, means our design's formulation does not rely on \textit{completely} avoiding (i.e., zero-forcing) the over-the-air self-interference channel or even the \textit{effective} self-interference channel, unlike several existing designs.
We assume we have channel knowledge of $\channel_{ii}$ but do not assume knowledge of $\channel_{ij}$ or $\channel_{ki}$, as was mentioned in \secref{sec:beams}.
We motivate this assumption by presuming that the self-interference channel can be more reliably estimated given its strength and can be done so possibly through calibration.
We assume large-scale quantities (e.g., transmit powers, large-scale channel gains, \glspl{snr}) are known.
Furthermore, our design holds for general symbol constellations, depending only on the symbol covariance and not the symbols themselves.


\subsection{Expressing Per-Antenna and Per-RF Chain Received Power Constraints}


%

	Let $\powersimaxlnawatts$ and $\powersimaxadcwatts$ be the maximum average self-interference power (in watts) over the symbol period allowed at each \lna and each \adc of the receiver of the full-duplex device $i$, respectively.
Referring to the terms presented in our system model and motivation, let us form our \lna and \adc constraints using $\powersimaxlnawatts$ and $\powersimaxadcwatts$.
A constraint bounding the symbol power\footnote{The term ``symbol power'' refers to the average power over the symbol period.} (in watts) at each \lna can be written as
\begin{gather}
\bars{\sqrt{\Ptxwatts\node{i}} G_{ii} \entry{\channel_{ii}}{\ell,:} \prerf\node{i} \prebb\node{i} \symbvec\node{i}}^2 \leq \powersimaxlnawatts 
\end{gather}
for all $\ell = 1, \dots, \Nr\node{i}$.
Collecting these $\Nr\node{i}$ constraints together, we can write
\begin{align}
\diag{\Ptxwatts\node{i} G_{ii}^2 \channel_{ii} \prerf\node{i} \prebb\node{i} \symbvec\node{i} \symbvec\ctransnode{i} \prebb\ctransnode{i} \prerf\ctransnode{i} \channel_{ii}\ctrans } \leq \powersimaxlnawatts \cdot \onevec{\Nr\node{i}} \label{eq:lna-constraint-1} 
\end{align}
where $\va \leq \vb$ denotes element-wise inequality. 
For a given channel realization, it is impractical to attempt to satisfy \eqref{eq:lna-constraint-1} on a per-symbol basis.
Furthermore, it may be computationally expensive, severely sub-optimal, or potentially impossible to ensure \eqref{eq:lna-constraint-1} is met for all symbol vectors in a constellation.
This motivates us to satisfy our constraint in expectation over $\symbvec\node{i}$ and, noting our defined symbol covariance \eqref{eq:symbol-covariance}, results in the constraint
\begin{gather}
\frac{\Ptxwatts\node{i} G_{ii}^2}{\Ns\node{i}} \cdot \diag{\channel_{ii} \prerf\node{i} \prebb\node{i} \prebb\ctransnode{i} \prerf\ctransnode{i} \channel_{ii}\ctrans } \leq \powersimaxlnawatts \cdot \onevec{\Nr\node{i}} \label{eq:lna-constraint-3}.
\end{gather}
In a similar fashion, by incorporating the analog combiner $\comrf\node{i}$, we can express our per-\rf chain received self-interference power constraint as
\begin{gather}
\frac{\Ptxwatts\node{i} G_{ii}^2}{\Ns\node{i}} \cdot \diag{\comrf\ctransnode{i} \channel_{ii} \prerf\node{i} \prebb\node{i} \prebb\ctransnode{i} \prerf\ctransnode{i} \channel_{ii}\ctrans \comrf\node{i}} \leq \powersimaxadcwatts \cdot \onevec{\Lr\node{i}} \label{eq:adc-constraint-3}
\end{gather}
which captures all $\Lr\node{i}$ \adcs in a single expression.

To abstract out the impact $\Ptxwatts\node{i}$ and $G_{ii}^2$ have on meeting our per-antenna power constraint $\powersimaxlnawatts$ and per-\rf chain power constraint $\powersimaxadcwatts$, we introduce the following unitless variables $\etalna$ and $\etaadc$, respectively, as
\begin{align}
\etalna
&\triangleq \frac{\powersimaxlnawatts}{\Ptxwatts\node{i} \cdot G_{ii}^2} 
, \quad \etaadc
\triangleq \frac{\powersimaxadcwatts}{\Ptxwatts\node{i} \cdot G_{ii}^2} \label{eq:eta-lna-adc-def}
\end{align}
which will provide more generalized analysis across combinations of $\powersimaxlnawatts$, $\powersimaxadcwatts$, $\Ptxwatts\node{i}$, and $G_{ii}^2$.
Note that stricter constraints are when $\etalna$ and $\etaadc$ are low (we permit little self-interference at the receiver), while relaxed constraints are when $\etalna$ and $\etaadc$ are high (we permit high self-interference at the receiver).
Using \eqref{eq:eta-lna-adc-def}, the constraints in \eqref{eq:lna-constraint-3} and \eqref{eq:adc-constraint-3} can be equivalently expressed as
\begin{gather}
\frac{1}{\Ns\node{i}} \cdot \svmaxsq{\channel_{ii} \prerf\node{i} \prebb\node{i}} \leq \etalna \label{eq:constraint-sv-lna} \\
\frac{1}{\Ns\node{i}} \cdot \svmaxsq{\comrf\ctransnode{i} \channel_{ii} \prerf\node{i} \prebb\node{i}} \leq \etaadc \label{eq:constraint-sv-adc}
\end{gather}
respectively, where $\svmax{\mA}$ denotes the maximum singular value of $\mA$.

\subsection{Satisfying our Constraints}

With our \lna and \adc constraints in hand, we turn our attention to producing a hybrid beamforming design that satisfies \eqref{eq:constraint-sv-lna} and \eqref{eq:constraint-sv-adc} while achieving an appreciable spectral efficiency on our two links.
The mutual information (under Gaussian signaling), or spectral efficiency (in bps/Hz), of the link from $i$ to $j$ is referred to as $\Rij$ and takes the familiar form in \eqref{eq:mi-ij} \cite{heath_lozano}.
Treating the effects of self-interference as noise, the mutual information of the link from $k$ to $i$ is referred to as $\Rki$ and expressed in \eqref{eq:mi-ki}.
We let $\mQ_{\mathrm{n}}\node{n}$ be the covariance of noise at the detector of $n \in \{i,j\}$ normalized to the noise power and let $\mQ_{\mathrm{int}}\node{i}$ be the covariance of self-interference at the detector of $i$ normalized to the noise power.
Since noise and the transmitted symbols from $i$ to $j$ are uncorrelated, the self-interference-plus-noise covariance can be written as $\mQ_{\mathrm{n}}\node{i} + \mQ_{\mathrm{int}}\node{i}$.
\begin{gather}
\mathcal{R}_{ij} = 
\logtwodet{\mI + \frac{\snr_{ij}}{\Ns\node{ij}} \combb\ctransnode{j} \comrf\ctransnode{j} \mH_{ij} \prerf\node{i} \prebb\node{i} \prebb\ctransnode{i} \prerf\ctransnode{i} \channel_{ij}\ctrans \comrf\node{j} \combb\node{j} \left(\mQ_{\mathrm{n}}\node{j}\right)\inv} \label{eq:mi-ij} \\
\mathcal{R}_{ki} = 
\logtwodet{ \mI + \frac{\snr_{ki}}{\Ns\node{ki}} \combb\ctransnode{i} \comrf\ctransnode{i} \mH_{ki} \prerf\node{k} \prebb\node{k} \prebb\ctransnode{k} \prerf\ctransnode{k} \channel_{ki}\ctrans \comrf\node{i} \combb\node{i} \left(\mQ_{\mathrm{n}}\node{i} + \mQ_{\mathrm{int}}\node{i} \right)\inv  } \label{eq:mi-ki}
\end{gather}


A sensible approach to design our system would be to maximize the sum spectral efficiency $\mathcal{R}_{ij} + \mathcal{R}_{ki}$ subject to our per-antenna and per-\rf chain self-interference power constraints as well as the precoding power constraint in \eqref{eq:baseband-precoder-power-constraint}, described below in problem \eqref{eq:problem-ideal-2}.
\begin{subequations} \label{eq:problem-ideal-2}
	\begin{align}
	&\max_{\substack{\parens{\prerf\node{i},\comrf\node{j}} \in \mathcal{T}_{ij} \\ \parens{\prerf\node{k},\comrf\node{i}} \in \mathcal{T}_{ki} }} \ 
	\max_{\substack{\prebb\node{i},\combb\node{j}\\\prebb\node{k},\combb\node{i}}} \ \Rij + \Rki \\
	& \ \ \st \ 
\frobeniustwo{\prebb\node{i}} \leq 1, \ \frobeniustwo{\prebb\node{k}} \leq 1 \label{eq:problem-ideal-2-constraint-power-2} \\
	& \ \ \qquad \frac{1}{\Ns\node{i}} \cdot \svmaxsq{\channel_{ii} \prerf\node{i} \prebb\node{i}} \leq \etalna \label{eq:problem-ideal-2-constraint-lna-1} \\
	& \ \qquad \frac{1}{\Ns\node{i}} \cdot \svmaxsq{\comrf\ctransnode{i} \channel_{ii} \prerf\node{i} \prebb\node{i}} \leq \etaadc \label{eq:problem-ideal-2-constraint-adc-1} 
	\end{align}
\end{subequations}
Note that we have restricted our choices for analog beamforming to the sets $\Tij$ and $\Tki$ supplied from beam candidate acquisition in \secref{sec:beams}.
Solving problem \eqref{eq:problem-ideal-2} is difficult for a variety of reasons, chiefly the non-convexity arising from the interplay of the precoder at $i$ in both transmit link performance and self-interference, meaning it impacts both $\Rij$ and $\Rki$.
This motivates us to split our design into two stages.
The first stage will be to configure a portion of our system subject to our constraints.
Then, with the constraints met by the first stage, the second stage of our design will configure the remaining precoders and combiners.

Let us begin the first stage of our design by defining $\mathcal{I}_{ij}$ as the mutual information afforded to the \rf chains of device $j$ by device $i$ as \eqref{eq:problem-obj}.
\begin{align}
\mathcal{I}_{ij} = \logtwodet{\mat{I} + \frac{\snr_{ij}}{\Ns\node{ij}} \comrf\ctransnode{j} \channel_{ij} \prerf\node{i} \prebb\node{i} \prebb\ctransnode{i} \prerf\ctransnode{i} \channel_{ij}\ctrans \comrf\node{j} (\comrf\ctransnode{j}\comrf\node{j})\inv } \label{eq:problem-obj}
\end{align}
Suppose during beam candidate acquisition, we let $\Kij = \Kki = 1$, leading to the analog beamformers $\prerf\node{i}$, $\comrf\node{j}$, $\prerf\node{k}$, and $\comrf\node{i}$ being fixed (i.e., the first and only candidates from $\Tij$ and $\Tki$).
In such a case, the responsibility to satisfy the per-antenna and per-\rf chain constraints lay solely in $\prebb\node{i}$ as evidenced by \eqref{eq:problem-ideal-2-constraint-lna-1}  and \eqref{eq:problem-ideal-2-constraint-adc-1}.
This leads us to formulate problem \eqref{eq:problem-inner-1}, where we aim to maximize this transmit link mutual information $\Iij$ subject to our per-antenna and per-\rf chain constraints and the aforementioned precoding power constraint in \eqref{eq:baseband-precoder-power-constraint}.
\begin{subequations} \label{eq:problem-inner-1}
\begin{align}
\max_{\prebb\node{i}} \ & \ \mathcal{I}_{ij} \label{eq:problem-inner-obj-1} \\
\st \ 
& \normfro{\prebb\node{i}}^2 \leq 1 \label{eq:problem-inner-constraint-power-1} \\
&\frac{1}{\Ns\node{i}} \cdot \svmaxsq{\channel_{ii} \prerf\node{i} \prebb\node{i}} \leq \etalna \label{eq:problem-inner-constraint-lna-1} \\
&\frac{1}{\Ns\node{i}} \cdot \svmaxsq{\comrf\ctransnode{i} \channel_{ii} \prerf\node{i} \prebb\node{i}} \leq \etaadc \label{eq:problem-inner-constraint-adc-1}
\end{align}
\end{subequations}
Solving problem \eqref{eq:problem-inner-1} would ensure that transmission from $i$ to $j$ is prioritized while preventing receiver-side saturation.
Problem \eqref{eq:problem-inner-1} is technically not convex but can be easily recast as such.

\begin{theorem}
Problem \eqref{eq:problem-inner-1} can be recast as a convex problem. 

\begin{proof}
Noting that problem \eqref{eq:problem-inner-1} is non-convex in $\prebb\node{i}$ but is convex in the product $\prebb\node{i} \prebb\ctransnode{i}$, we rewrite problem \eqref{eq:problem-inner-1} using the substitution $\xbb\node{i} = \prebb\node{i} \prebb\ctransnode{i} \ispsd \mat{0}$ as
\begin{subequations} \label{eq:problem-inner-2}
	\begin{align}
	\max_{\xbb\node{i}\ispsd \mat{0}} \ & \ \Iij \label{eq:problem-inner-obj-2} \\
	\st \ 
	& \normfro{\prebb\node{i}}^2 \leq 1 \label{eq:problem-inner-constraint-power-2} \\
	& \frac{1}{\Ns\node{i}} \cdot \svmaxsq{\channel_{ii} \prerf\node{i} \prebb\node{i}} \leq \etalna \label{eq:problem-inner-constraint-lna-2} \\
	& \frac{1}{\Ns\node{i}} \cdot \svmaxsq{\comrf\ctransnode{i} \channel_{ii} \prerf\node{i} \prebb\node{i}} \leq \etaadc \label{eq:problem-inner-constraint-adc-2} \\
	& \prebb\node{i} \prebb\ctransnode{i} = \xbb\node{i}
	\end{align}
\end{subequations}
With this simple restructuring, problem \eqref{eq:problem-inner-2} is convex and can be solved efficiently using a convex solver (e.g., CVX \cite{cvx} was used to evaluate our design in \secref{sec:simulation-results}).
Once solved, $\xbb\node{i}$ can be factored to recover $\prebb\node{i}$. 
\end{proof}

\begin{remark}
The solution to problem \eqref{eq:problem-inner-2} is unique but the factorization $\xbb\node{i} = \prebb\node{i} \prebb\ctransnode{i}$ is not.
As such, our problem is a function of the digital precoder covariance and not of the digital precoder itself.
In other words, the solution $\xbb\node{i}$ to problem \eqref{eq:problem-inner-2} can be factored arbitrarily, since $\Lt\node{i} = \Ns\node{ij}$, to retrieve a globally optimal $\prebb\node{i}$.
\end{remark}
\begin{remark}
Since $\xbb\node{i} = \prebb\node{i} \prebb\ctransnode{i} = \mat{0}$ is always a solution to problem \eqref{eq:problem-inner-2}, it is feasible.
Intuitively, this can be attributed to the fact that it is always possible for the transmitter at $i$ to shut off completely to ensure the \lnas and \adcs do not saturate.
\end{remark}
\begin{remark}
At least one of the constraints \eqref{eq:problem-inner-constraint-power-2}--\eqref{eq:problem-inner-constraint-adc-2} will be tight (have equality) under the optimal solution to problem \eqref{eq:problem-inner-2}. 
Intuitively, this can be attributed to the fact that additional power should be supplied to the digital precoder $\prebb\node{i}$---which will increase the mutual information $\Iij$---until it exceeds its power budget or self-interference power is too high at an \lna or \adc.
\end{remark}
\end{theorem}

To offer the system more freedom in its design, we now incorporate our sets of candidate beams $\Tij$ and $\Tki$ when $\Kij, \Kki \geq 1$.
By doing so, the diversity between candidate beams may allow our full-duplex transceiver to better transmit while meeting the constraints.
We capture this freedom in choosing our analog beamformers by wrapping problem \eqref{eq:problem-inner-1} with an outer maximization over $\Tij$ and $\Tki$, resulting in the following optimization problem.
\begin{subequations} \label{eq:problem-full-1}
	\begin{align}
	& \max_{\substack{\parens{\prerf\node{i},\comrf\node{j}} \in \mathcal{T}_{ij} \\ \parens{\prerf\node{k},\comrf\node{i}} \in \mathcal{T}_{ki} }} \ 
	\max_{\prebb\node{i}} \ \mathcal{I}_{ij} \label{eq:problem-full-obj-1} \\
	& \ \ \st \  \frobeniustwo{\prebb\node{i}} \leq 1 \label{eq:problem-full-constraint-power-1} \\
	& \qquad \frac{1}{\Ns\node{i}} \cdot \svmaxsq{\channel_{ii} \prerf\node{i} \prebb\node{i}} \leq \etalna \label{eq:problem-full-constraint-lna-1} \\
	& \qquad \frac{1}{\Ns\node{i}} \cdot \svmaxsq{\comrf\ctransnode{i} \channel_{ii} \prerf\node{i} \prebb\node{i}} \leq \etaadc \label{eq:problem-full-constraint-adc-1}
	\end{align}
\end{subequations}
Having shown that the inner maximization can be solved via the convex reformulation in \eqref{eq:problem-inner-2}, we can solve problem \eqref{eq:problem-full-1} exhaustively over all possible candidate combinations in $\mathcal{T}_{ij}$ and $\mathcal{T}_{ki}$.
Thus, \eqref{eq:problem-full-1} can be solved by solving the inner maximization $\Kij \times \Kki$ times.

Note that, as evidenced in \eqref{eq:problem-obj}, $\Iij$ contains $\channel_{ij}$, which we have not assumed explicit knowledge of.
Instead, using $\mathcal{H}_{ij}$ from \eqref{eq:rf-cand-channels-ij}, we do have knowledge of the effective channel $\comrf\ctransnode{j} \channel_{ij} \prerf\node{i}$ for all $(\prerf\node{i},\comrf\node{i}) \in \mathcal{T}_{ij}$, which can be used when solving problem \eqref{eq:problem-full-1}.
Furthermore, note that we do not require knowledge of $\channel_{ki}$ or $\comrf\ctransnode{i} \channel_{ki} \prerf\node{k}$ to solve \eqref{eq:problem-full-1}.
We do require knowledge of $\channel_{ii}$ to construct constraints \eqref{eq:problem-full-constraint-lna-1} and \eqref{eq:problem-full-constraint-adc-1}, which we have assumed knowledge of. 
Notice, however, that perhaps an estimate of $\comrf\ctransnode{i} \channel_{ii} \prerf\node{i}$ can be used to compute the per-\rf chain constraint \eqref{eq:problem-full-constraint-adc-1} since it may be estimated more reliably and frequently than that of $\channel_{ii}$, given its relatively small size and fully-digital nature.

Solving problem \eqref{eq:problem-full-1} yields the design for five of the eight precoding and combining matrices: $\combb\node{j}$, $\prebb\node{k}$, and $\combb\node{i}$ remain to be designed.
Designing these will take place in the next stage of our design.
Having prevented receiver-side components from saturating with appropriately chosen $\etalna$ and $\etaadc$, the receive chain at device $i$ is approximately linear, allowing for a much more straightforward design of the receive link and preventing severe degradation of as desired receive signal.

\subsection{Constraint Redundancy Conditions}
We now pause from our design to more closely examine the interplay between our precoding power constraint, per-antenna self-interference power constraint, and per-\rf chain self-interference power constraint.
In doing so, we will see that under appropriate conditions the latter two constraints may be inherently met by the precoding power constraint and under other conditions, the per-\rf chain self-interference power constraint may be inherently met by the per-antenna self-interference power constraint.
Knowledge of these conditions can accelerate solving problem \eqref{eq:problem-full-1}.

\begin{theorem} \label{thm:lna-redundant-by-power}
When condition \eqref{eq:constraint-lna-met-by-power} holds, the per-antenna constraint \eqref{eq:problem-full-constraint-lna-1} is implicitly met by the precoding power constraint \eqref{eq:problem-full-constraint-power-1}.
\begin{align}
\etalna \geq \frac{1}{\Ns\node{i}} \cdot \svmaxsq{\channel_{ii} \prerf\node{i}} \label{eq:constraint-lna-met-by-power}
\end{align}
\begin{proof}
	Since the precoding power constraint \eqref{eq:problem-inner-constraint-power-1} is satisfied, we note that 
	\begin{align}
	\svmaxsq{\prebb\node{i}} \leq \normfro{\prebb\node{i}}^2 \leq 1 \label{eq:power-satisfied-sv}
	\end{align}
	which allows us to see that
	\begin{align}
	\etalna 
	&\geq \frac{1}{\Ns\node{i}} \cdot \svmaxsq{\channel_{ii} \prerf\node{i}} \ \\
	&\geq \frac{1}{\Ns\node{i}} \cdot \svmaxsq{\channel_{ii} \prerf\node{i}} \cdot \svmaxsq{\prebb\node{i}}  \\
	&\geq \frac{1}{\Ns\node{i}} \cdot \svmaxsq{\channel_{ii} \prerf\node{i} \prebb\node{i}}
	\end{align}
	where we have used $\svmaxsq{\mA\mB} \leq \svmaxsq{\mA} \cdot \svmaxsq{\mB}$.
\end{proof}
\end{theorem}


\begin{theorem}
When condition \eqref{eq:constraint-adc-redundant} holds, the per-\rf chain constraint \eqref{eq:problem-full-constraint-adc-1} is implicitly met by the precoding power constraint \eqref{eq:problem-full-constraint-power-1}.
\begin{align}
\etaadc 
&\geq \frac{1}{\Ns\node{i}} \cdot \svmaxsq{\comrf\ctransnode{i} \channel_{ii} \prerf\node{i}} \label{eq:constraint-adc-redundant}
\end{align}
\begin{proof}
Using the fact from \eqref{eq:power-satisfied-sv} and assuming \eqref{eq:constraint-adc-redundant} to be true, we can write
\begin{align}
\etaadc 
&\geq \frac{1}{\Ns\node{i}} \cdot \svmaxsq{\comrf\ctransnode{i} \channel_{ii} \prerf\node{i}} \\
&\geq \frac{1}{\Ns\node{i}} \cdot \svmaxsq{\comrf\ctransnode{i} \channel_{ii} \prerf\node{i}} \cdot \svmaxsq{\prebb\node{i}}  \\
&\geq \frac{1}{\Ns\node{i}} \cdot \svmaxsq{\comrf\ctransnode{i} \channel_{ii} \prerf\node{i} \prebb\node{i}}
\end{align}
\end{proof}
\end{theorem}

\begin{remark}
When \eqref{eq:constraint-lna-met-by-power} and \eqref{eq:constraint-adc-redundant} hold, the (half-duplex) capacity-achieving strategy on the transmit link is the solution to problem \eqref{eq:problem-inner-1} since the per-antenna and per-\rf chain constraints vanish, leaving only the precoding power constraint.
\end{remark}

\begin{theorem} \label{thm:adc-redundant-by-lna}
When the per-antenna constraint \eqref{eq:problem-full-constraint-lna-1} is satisfied and condition \eqref{eq:constraint-adc-redundant-by-lna} holds, the per-\rf chain constraint \eqref{eq:problem-full-constraint-adc-1} is satisfied.
\begin{align}
\etaadc \geq \etalna \cdot \svmaxsq{\comrf\ctransnode{i}} \label{eq:constraint-adc-redundant-by-lna}
\end{align}
\begin{proof}
Starting with our assumption that the per-antenna constraint \eqref{eq:problem-full-constraint-lna-1} is satisfied, we have
\begin{align} 
\frac{1}{\Ns\node{i}} \cdot \svmaxsq{\channel_{ii} \prerf\node{i} \prebb\node{i}}
& \leq \etalna 
 \leq \frac{\etaadc}{\svmaxsq{\comrf\ctransnode{i}}}
\end{align}
which yields
\begin{align} 
\etaadc 
& \geq \frac{1}{\Ns\node{i}} \cdot \svmaxsq{\comrf\ctransnode{i}} \cdot \svmaxsq{\channel_{ii} \prerf\node{i} \prebb\node{i}} \\
& \geq \frac{1}{\Ns\node{i}} \cdot \svmaxsq{\comrf\ctransnode{i} \channel_{ii} \prerf\node{i} \prebb\node{i}}
\end{align}
which indicates directly that \eqref{eq:problem-full-constraint-adc-1} is satisfied.
\end{proof}

\begin{corollary}
	When $\etalna = 0$ and $\etaadc > 0$, satisfying the per-antenna constraint \eqref{eq:problem-full-constraint-lna-1} also satisfies the per-\rf chain constraint \eqref{eq:problem-full-constraint-adc-1}. 
\end{corollary}

\begin{remark}
A condition analogous to \thmref{thm:adc-redundant-by-lna} stating that satisfying the per-antenna constraint \eqref{eq:problem-full-constraint-lna-1} based solely on meeting the per-\rf chain constraint \eqref{eq:problem-full-constraint-adc-1} is not possible.
\end{remark}
\end{theorem}

\subsection{Remainder of the Design}
We now complete our \mmwave \mimo design.
Solving \eqref{eq:problem-full-1} yields selections for $\prebb\node{i}$, $\prerf\node{i}$, $\comrf\node{j}$, $\prerf\node{k}$, and $\comrf\node{i}$.
Configuring $\combb\node{j}$, $\prebb\node{k}$, and $\combb\node{i}$ remains, which we execute as follows.
Having the rest of the transmit link configured, the optimal linear baseband combiner at $j$ can be designed in a \lmmse fashion as follows.
Let $\tilde{\channel}_{ij} \triangleq \comrf\ctransnode{j} \channel_{ij} \prerf\node{i} \prebb\node{i} \in  \setmatrixcomplex{\Lr\node{j}}{\Ns\node{ij}}$ be the effective transmit channel after solving \eqref{eq:problem-full-1}.
Note that this can computed from the product of $\comrf\ctransnode{j} \channel_{ij} \prerf\node{i}$, which is referenced from $\mathcal{H}_{ij}$, and $\prebb\node{i}$.
Then, the \lmmse baseband combiner at $j$ can be constructed as
\begin{align}
\combb\node{j} 
&= \frac{1}{\sqrt{\Ptx\node{i}} G_{ij}} \parens{ \tilde{\channel}_{ij} \tilde{\channel}_{ij}\ctrans + \frac{\Ns\node{ij}}{\snr_{ij}} \comrf\ctransnode{j} \comrf\node{j} }\inv \tilde{\channel}_{ij}.
\end{align}
Since $\prebb\node{i}$ will not generally diagonalize the effective channel (due to meeting the per-antenna and per-\rf chain constraints), an \lmmse combiner at $j$ will aim to reduce inter-stream interference and reject noise.
This concludes configuration of the transmit link, having set $\prebb\node{i}$, $\prerf\node{i}$, $\comrf\node{j}$, and $\combb\node{j}$.

We now turn our attention to the receive link, where we need to configure $\prebb\node{k}$ and $\combb\node{i}$.
Now that we have made our selections of $\prerf\node{k}$ and $\comrf\node{i}$, we begin by estimating the relatively small channel $\tilde{\channel}_{ki} \triangleq \comrf\ctransnode{i} \channel_{ki} \prerf\node{k} \in \setmatrixcomplex{\Lr\node{i}}{\Lt\node{k}}$, 
which can be observed digitally and we assume is error-free.
Taking the \gls{svd} of this effective channel from $k$ to $i$ and accounting for noise coloring, we get $\mU_{ki} \mSigma_{ki} \mV_{ki}\ctrans = \svdop{\parens{\comrf\ctransnode{i} \comrf\node{i}}^{-1/2} \tilde{\channel}_{ki}}$.
We then build the precoder that maximizes the mutual information offered to the \rf chains of $i$ as
\begin{align}
\prebb\node{k} = \entry{\mV_{ki}}{:,1:\Ns\node{ki}} \times \mP\node{k}
\end{align}
where $\mP\node{k}$ is a diagonal water-filling power allocation matrix \cite{heath_lozano}. 

Finally, we are left to configure $\combb\node{i}$.
Before doing so, however, we notice that $\combb\node{i}$ and the received symbols it acts on exist in the digital domain, residing after the \adcs.
Having knowledge of $\channel_{ii}$, $\symbvec\node{i}$, and all other beamformers, we can synthesize the received self-interference and subtract it before applying our combiner $\combb\node{i}$.
Recall that the symbol vector after the \adcs is
\begin{align}
\ydig 
&= \yadcdes + \yadcint + \yadcnoise + \equant.
\end{align}
Since $\ydig$ is in the digital domain, we can compute $\yadcint$ perfectly as
\begin{align}
\yadcint = \sqrt{\Ptx\node{i}} G_{ii} \comrf\ctransnode{i} \channel_{ii} \prerf\node{i} \prebb\node{i} \symbvec\node{i}
\end{align}
and can subtract it from $\ydig$ before applying our combiner.
Subtracting self-interference from $\ydig$, we get
\begin{align}
\vy\node{i} 
&= \ydig - \yadcint 
= \yadcdes + \yadcnoise + \equant \label{eq:dsic-post}.
\end{align}
We can estimate the symbols intended for $i$ from $k$ by applying a combiner $\combb\node{i}$ to $\vy\node{i}$ as
\begin{align}
\symbvecest\node{k} 
&= \combb\ctransnode{i} \vy\node{i} = \combb\ctransnode{i} \parens{\yadcdes + \yadcnoise + \equant} \label{eq:combiner-additive-noise} 
\end{align}
which will be corrupted by additive noise and the effects of quantization.
At first glance, it appears that self-interference does not play a role our symbol estimate $\symbvecest\node{k}$. 
However, given that our \adcs have limited resolution, the power of the quantization error term $\equant$ will increase with increased self-interference power.
Recall that this is precisely the motivation for our per-\rf chain self-interference power constraint.

As evidenced by \eqref{eq:combiner-additive-noise}, we can see that the linear combiner $\combb\node{i}$ acts on a desired signal plus two noise terms.
Before proceeding, let us find the covariance of $\equant$.
Finding the covariance of the symbols reaching the \adcs, we get
\begin{align}
\ev{\yadc\yadc\ctrans} =
& \ \frac{\Ptx\node{k} G_{ki}^2}{\Ns\node{ki}} \cdot \comrf\ctransnode{i} \channel_{ki} \prerf\node{k} \prebb\node{k} \prebb\ctransnode{k} \prerf\ctransnode{k} \channel_{ki}\ctrans \comrf\node{i} \nonumber \\
& \ + \frac{\Ptx\node{i} G_{ii}^2}{\Ns\node{ij}} \cdot \comrf\ctransnode{i} \channel_{ii} \prerf\node{i} \prebb\node{i}  \prebb\ctransnode{i} \prerf\ctransnode{i} \channel_{ii}\ctrans \comrf\node{i} \nonumber + \noisevar \cdot \comrf\ctransnode{i} \comrf\node{i} \nonumber.
\end{align}
Applying \eqref{eq:power-quant-1} per-\adc allows us to write the covariance matrix of $\equant$ as
\begin{align}
\mR_{\mathrm{quant}}
&= \ev{\equant \equant\ctrans} 
= \frac{8}{12 \cdot 2^{2b}} \cdot \mI \had \ev{\yadc\yadc\ctrans}
\end{align}
where $\had$ denotes the Hadamard (element-wise) product.
The linear design of $\combb\node{i}$ that minimizes the \gls{mse} of $\symbvecest\node{k}$ is 
\begin{align} \label{eq:mmse-i}
\combb\node{i}
&= \frac{1}{\sqrt{\Ptx\node{k}} G_{ki}} \parens{ \tilde{\channel}_{ki} \tilde{\channel}_{ki}\ctrans + \frac{\Ns\node{ki}}{\snr_{ki}} \comrf\ctransnode{i} \comrf\node{i} + \frac{\Ns\node{ki}}{\Ptx\node{k} G_{ki}^2} \mR_{\mathrm{quant}} }\inv \tilde{\channel}_{ki}.
\end{align}

Now that our design is complete, we characterize the covariance of self-interference and of noise at the detector, which can be used to evaluate $\mathcal{R}_{ij}$ and $\mathcal{R}_{ki}$ in \eqref{eq:mi-ij} and \eqref{eq:mi-ki}, respectively.
Let $\mQ_{\mathrm{int}}\node{n}$---the covariance of the received quantization noise due to self-interference at $i$, normalized to the noise power---be $\mQ_{\mathrm{int}}\node{i} = \frac{1}{\noisevar} \combb\ctransnode{i} \mR_{\mathrm{quant}} \combb\node{i}$.
Let $\mQ_{\mathrm{n}}\node{n}$---the covariance of the received noise at $n \in \{i,j\}$, normalized to the noise power---be written as $\mQ_{\mathrm{n}}\node{n} = \combb\ctransnode{n} \comrf\ctransnode{n} \comrf\node{n} \combb\node{n}$.
This concludes our design, which we evaluate in the following section.











\section{Numerical Results} \label{sec:simulation-results}




We have simulated our system in a Monte Carlo fashion with the following parameters.
For simplicity, we use $32$-element, half-wavelength uniform linear arrays with isotropic elements at all devices, where the horizontal transmit and receive array at $i$ are separated vertically by $10$ wavelengths.
Each transmitter and receiver is equipped with $2$ \rf chains and multiplexes $2$ streams.
The transmit power at each device is $30$ dBm, while the noise power is $-85$ dBm.

To model the channel between the transmit and receive arrays of device $i$, we use the following summation, which was originally suggested in \cite{xia_2017,satyanarayana_hybrid_2019},
\begin{align} \label{eq:si-channel-rice}
\channel_{ii} &= \sqrt{\frac{\kappa}{\kappa + 1}} \channel^{\mathrm{NF}}_{ii} + \sqrt{\frac{1}{\kappa + 1}} \channel^{\mathrm{FF}}_{ii}
\end{align}
where the Rician factor $\kappa$ captures the amount of power in the near-field portion relative to the far-field portion.
The near-field component is modeled using a spherical-wave model \cite{spherical_2005} as $\left[\channel^{\textrm{NF}}_{ii}\right]_{v,u} = \frac{\gamma}{r_{u,v}}\exp \left(-\j 2 \pi \frac{r_{u,v}}{\lambda} \right)$, 
where $r_{u,v}$ is the distance between the $u$-th transmit antenna and the $v$-th receive antenna, $\lambda$ is the carrier wavelength, and $\gamma$ ensures that the channel is normalized such that $\ev{\frobeniustwo{\channel_{ii}}} = \Nt\node{i}\Nr\node{i}$. 
Note that this near-field model is deterministic for a given relative array geometry at $i$.
The far-field component captures reflections from the environment and is modeled using \eqref{eq:desired-channel} with $\numrays \sim \distuniform{1}{15}$.
The transmit and receive channels are modeled with $\numrays \sim \distuniform{4}{15}$.
For both channel models, each ray's \gls{aod} and \gls{aoa} are drawn from $\distuniform{-\pi/2}{\pi/2}$.
During beam candidate acquisition, we assume properly normalized \gls{dft} codebooks are used.
Furthermore, we assume the number of measurements taken during beam candidate acquisition (i.e., $\Mt\node{i}$, $\Mr\node{j}$, $\Mt\node{k}$, and $\Mr\node{i}$) is sufficiently large on each link such that $\Tij$ and $\Tki$ are built using the strongest rays in their respective channels.

	Let us define the transmit link capacity $\Cij$ as the maximum spectral efficiency possible on the link from $i$ to $j$ when drawing the analog precoder $\prerf\node{i}$ and analog combiner $\comrf\node{j}$ from the beam candidate set $\Tij$ as 
	\begin{gather} \label{eq:capacity-ij-full}
	\mathcal{C}_{ij} = 
	\max_{\substack{\prebb\node{i}, \parens{\prerf\node{i},\comrf\node{j}} \in \mathcal{T}_{ij}}} 
	\ \mathcal{I}_{ij} \quad
	\st \ 
	\frobeniustwo{\prebb\node{i}} \leq 1
	\end{gather}
	which can be achieved using the well known method of water-filled eigenbeamforming.
	Let us define the receive link capacity $\Cki$ as the maximum spectral efficiency possible on the link from $k$ to $i$ when drawing the analog precoder $\prerf\node{k}$ and the analog combiner $\comrf\node{i}$ from the candidate set $\Tki$ as
	\begin{gather} \label{eq:capacity-ki-full}
	\mathcal{C}_{ki} = 
	\max_{\substack{\prebb\node{k}, \parens{\prerf\node{k},\comrf\node{i}} \in \mathcal{T}_{ki}}} 
	\ \mathcal{I}_{ki} \quad
	\st \ 
	\frobeniustwo{\prebb\node{k}} \leq 1
	\end{gather}
	where
	\begin{align} \label{eq:Iki}
	\mathcal{I}_{ki} = \logtwodet{ \mI + \frac{\snr_{ki}}{\Ns\node{ki}} \comrf\ctransnode{i} \mH_{ki} \prerf\node{k} \prebb\node{k} \prebb\ctransnode{k} \prerf\ctransnode{k} \channel_{ki}\ctrans \comrf\node{i} \left( \comrf\ctransnode{i} \comrf\node{i} \right)\inv }.
	\end{align}	

While $\Cij$ and $\Cki$ are not the true channel capacities of $\channel_{ij}$ and $\channel_{ki}$, it is more meaningful when evaluating our results to use our codebook-based analog beamforming approach to accurately interpret the spectral efficiency gains (and costs) associated with our design versus a half-duplex system that is offered the same freedom in analog beamforming contained in $\Tij$ and $\Tki$.
Our design will, therefore, hope to achieve a sum spectral efficiency $\Rij + \Rki \geq \max\braces{\Cij,\Cki}$ to justify operating in a full-duplex fashion rather than a half-duplex one.



\begin{figure}[!t]
	\centering
	\includegraphics[width=0.47\linewidth,height=0.4\textheight,keepaspectratio]{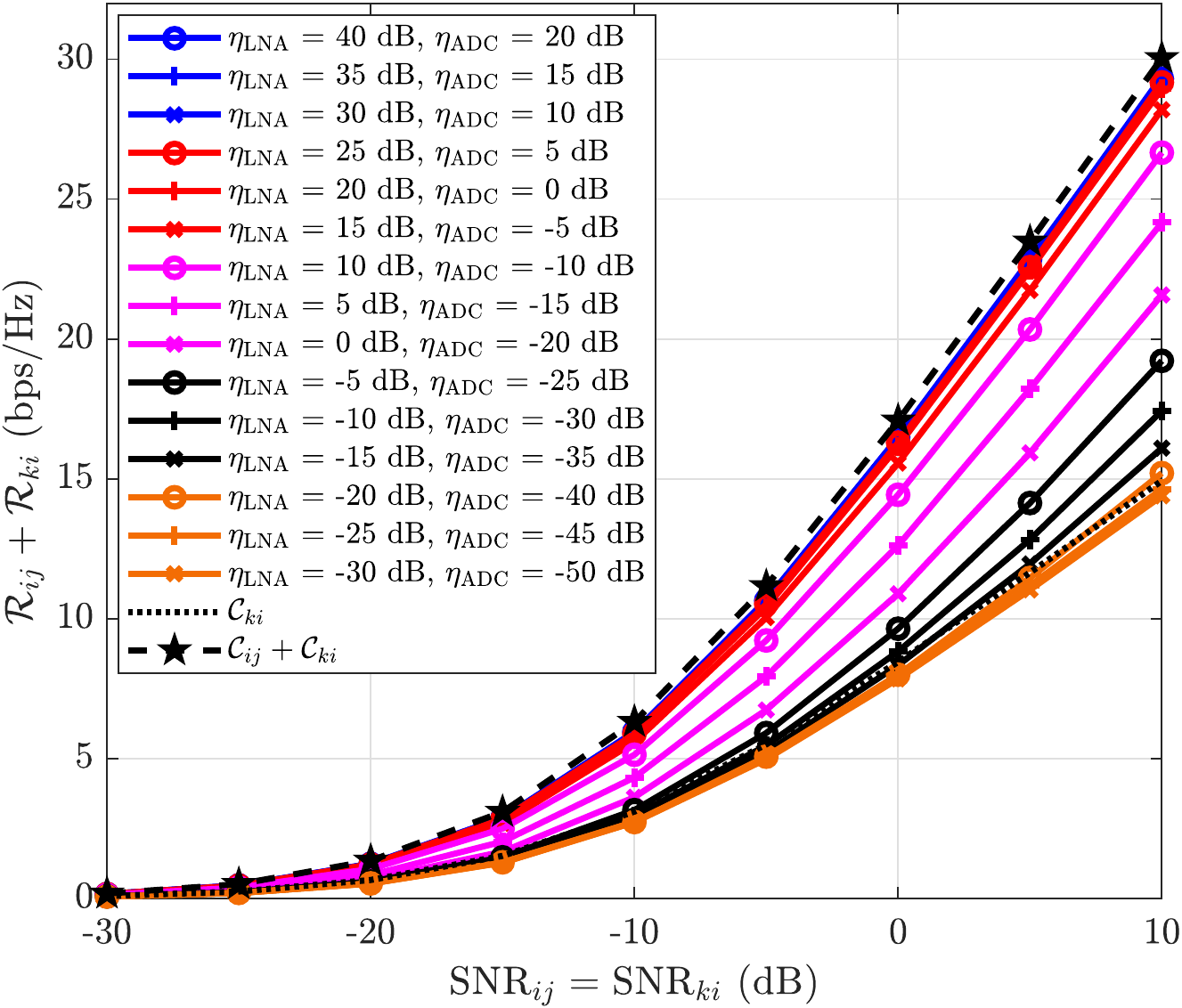}
	\caption{Sum spectral efficiency as a function of \gls{snr} for various $\etalna = \etaadc + 20$ dB, where $\kappa = 10$ dB, $\adcbits = 12$ bits, and $K_{ij} = K_{ki} = 3$. While not obvious from this figure, $\Rki$ is fairly robust to self-interference with $\adcbits = 12$ bits. Stricter choices of $\etalna = \etaadc + 20$ dB degrade $\Rij$ and, thus, the sum $\Rij + \Rki$.}
	\label{fig:results-se-snr}
\end{figure}

As intuition suggests, \textbf{the stricter the \lna and \adc constraints, the greater the sacrifice made on the transmit link's spectral efficiency to meet these constraints}.
\figref{fig:results-se-snr} confirms this, where the sum spectral efficiency $\Rij + \Rki$ as a function of \gls{snr} is evaluated at various \lna and \adc power constraints.
While not explicitly shown, the loss in sum spectral efficiency due to decreasing $\etalna = \etaadc + 20$ dB is due to loss in $\Rij$ in its attempt to prevent receiver-side saturation.
Given that we are primarily preserving the receive link by limiting the self-interference power reaching it, the receive link sees little sacrifice as a function of $\etalna = \etaadc + 20$ dB, having assumed the resolution of the \adc is $b = 12$ bits, which is fairly robust to these levels of $\etalna = \etaadc + 20$ dB (more on this later).
Therefore, the lower bound on $\Rij + \Rki$, as one would hope, is approximately the half-duplex receive capacity $\Cki$.
The small gap when $\Rij + \Rki < \Cki$ at very low $\etalna = \etaadc + 20$ dB in \figref{fig:results-se-snr} can be attributed to a small degree of quantization noise and, more significantly, the fact that the analog beamforming candidate chosen from $\Tki$ is not always the $\Cki$-achieving one, given that we have $\Kki = 3$.
In other words, by design, the candidate from $\Tki$ that maximizes transmit link performance subject to our constraints may not be the one that maximizes receive link performance.

\begin{figure*}[!t]
	\centering
	\subfloat[Transmit link spectral efficiency.]{\includegraphics[width=0.47\linewidth,height=0.4\textheight,keepaspectratio]{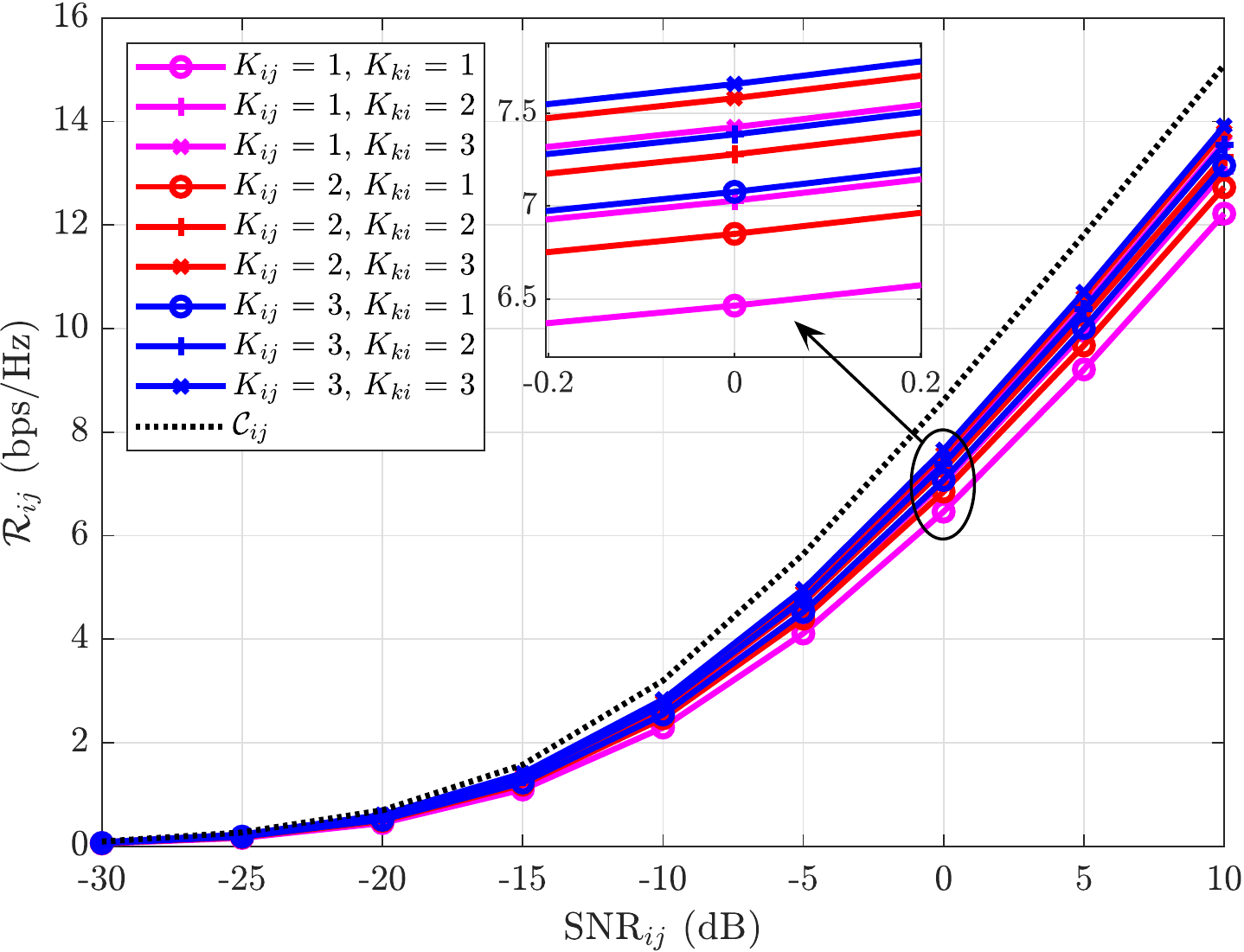}
		\label{fig:results-cand-snr-a}}
	\quad
	\subfloat[Sum spectral efficiency.]{\includegraphics[width=0.47\linewidth,height=0.4\textheight,keepaspectratio]{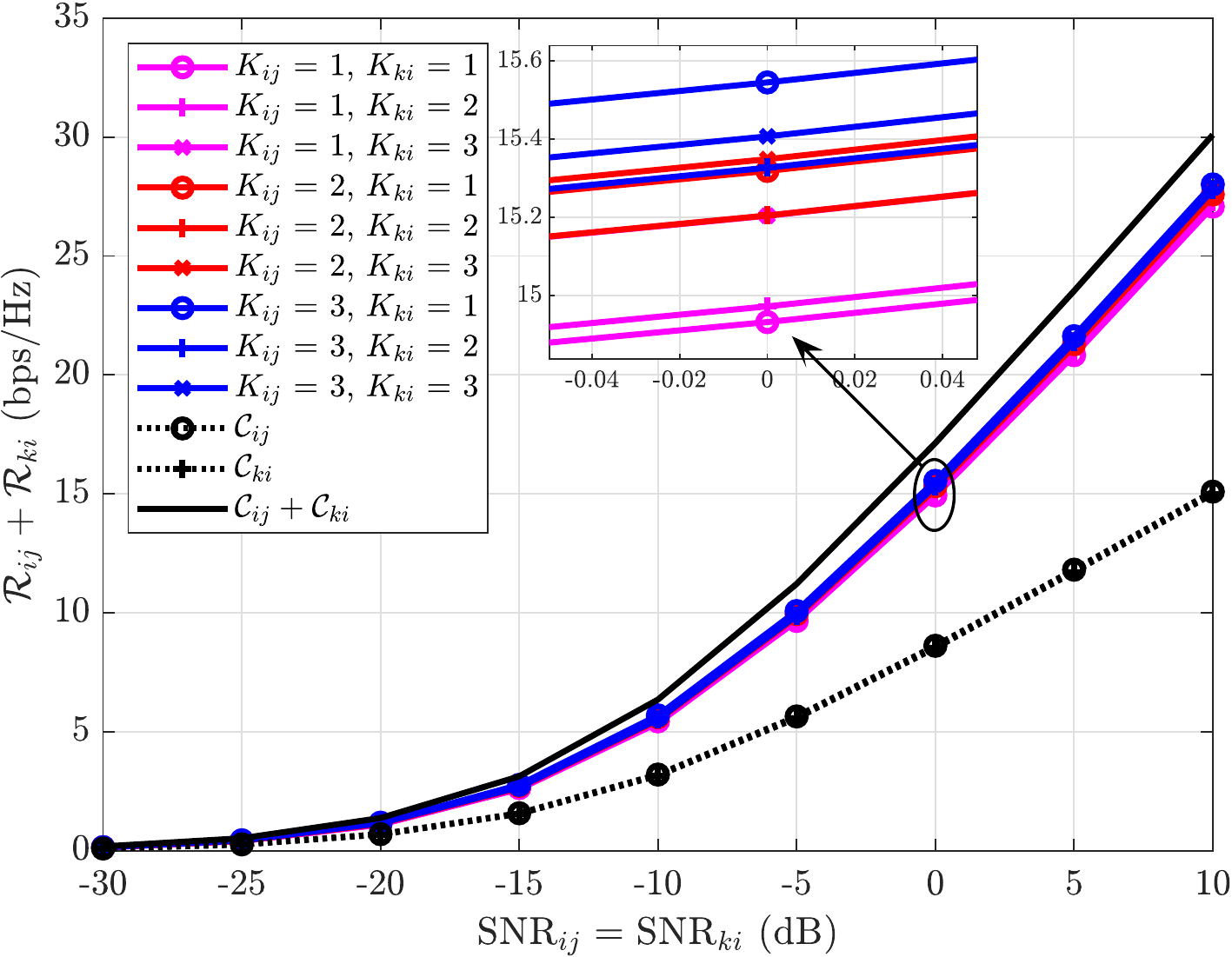}
		\label{fig:results-cand-snr-b}}
	\caption{Spectral efficiency as a function of \gls{snr} for various $\Kij$ and $\Kki$, where $\kappa = 10$ dB, $\etalna = 15$ dB, $\etaadc = -5$ dB, and $\adcbits = 12$ bits. Increasing $\Kij$ and $\Kki$ can only improve $\Rij$, whereas increasing $\Kki$ may degrade $\Rki$. The choice of $\Kij$ and $\Kki$ that maximizes the $\Rij + \Rki$ varies with each realization.}
	\label{fig:results-cand-snr}
\end{figure*}

\figref{fig:results-cand-snr} exhibits the gains in spectral efficiency afforded by increasing the number of analog beamforming candidates to our design.
\textbf{By increasing $\Kij$ and $\Kki$, the system can improve its performance on the \textit{transmit} link while meeting the per-antenna and per-\rf chain constraints.}
Referring to \figref{fig:results-cand-snr-a}, when $\Kij$ and $\Kki$ increases from having only one candidate (i.e., $K_{ij} = K_{ki} = 1$) to having three candidates on each link (i.e., $K_{ij} = K_{ki} = 3$), we see a gain of approximately 1.25 bps/Hz in $\Rij$ on average.
This can be attributed to the fact that widening the search space will yield greater flexibility in meeting the constraints while maximizing performance on the transmit link.
That is, rather than our optimization problem taking place over only $\prebb\node{i}$, it also takes place over the candidates in $\Tij$ and $\Tki$.
Interestingly, we can see that, on average, supplying our design with increased $K_{ki}$ has a relatively greater impact than $K_{ij}$; this can be seen by the fact that $\Kij = 1, \Kki = 3$ outperforms $\Kij = 3, \Kki = 1$ and even $\Kij = 3, \Kki = 2$ in terms of $\Rij + \Rki$.
With increased $\Kki$, some sacrifices may be made on the receive link by choosing a candidate from $\Tki$ that is not the $\Cki$-achieving one.
In general, $\Kij$ and $\Kki$ can be chosen to throttle performance between the transmit link and receive link.
Choosing a small $\Kki$, for example, preserves the receive link but reduces the full-duplex device's flexibility in avoiding self-interference.
For a fixed $\Kki$, choosing a large $\Kij$ can generally only help the system by increasing $\Rij$, though, increasing the number of candidates ($\Kij$ or $\Kki$) adds to the overhead associated with our design.


\comment{
\begin{figure*}[!t]
	\centering
	\subfloat[Transmit link spectral efficiency.]{\includegraphics[width=0.465\linewidth,height=0.4\textheight,keepaspectratio]{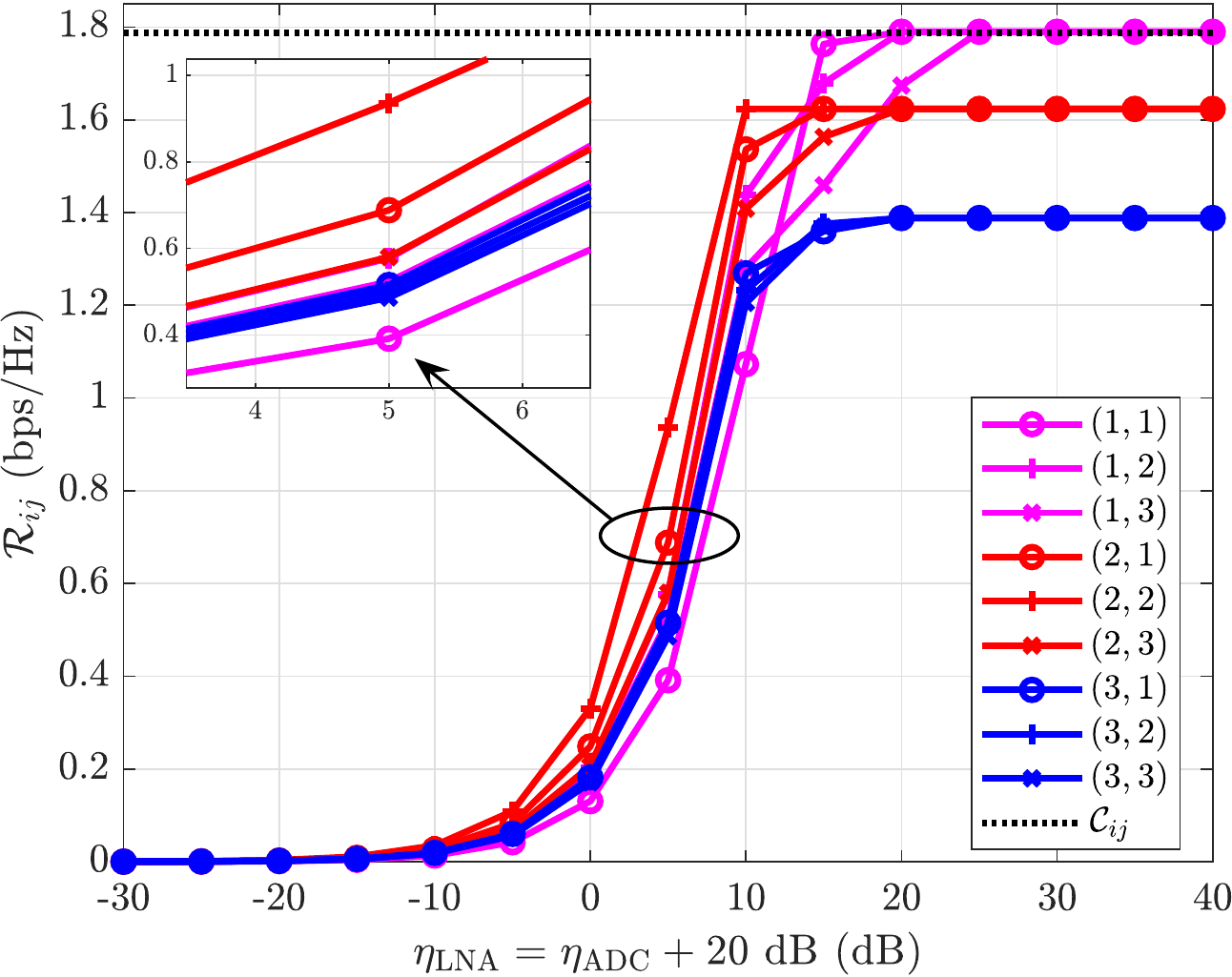}
		\label{fig:results-cand-eta-a}}
	\quad
	\subfloat[Sum spectral efficiency.]{\includegraphics[width=0.465\linewidth,height=0.4\textheight,keepaspectratio]{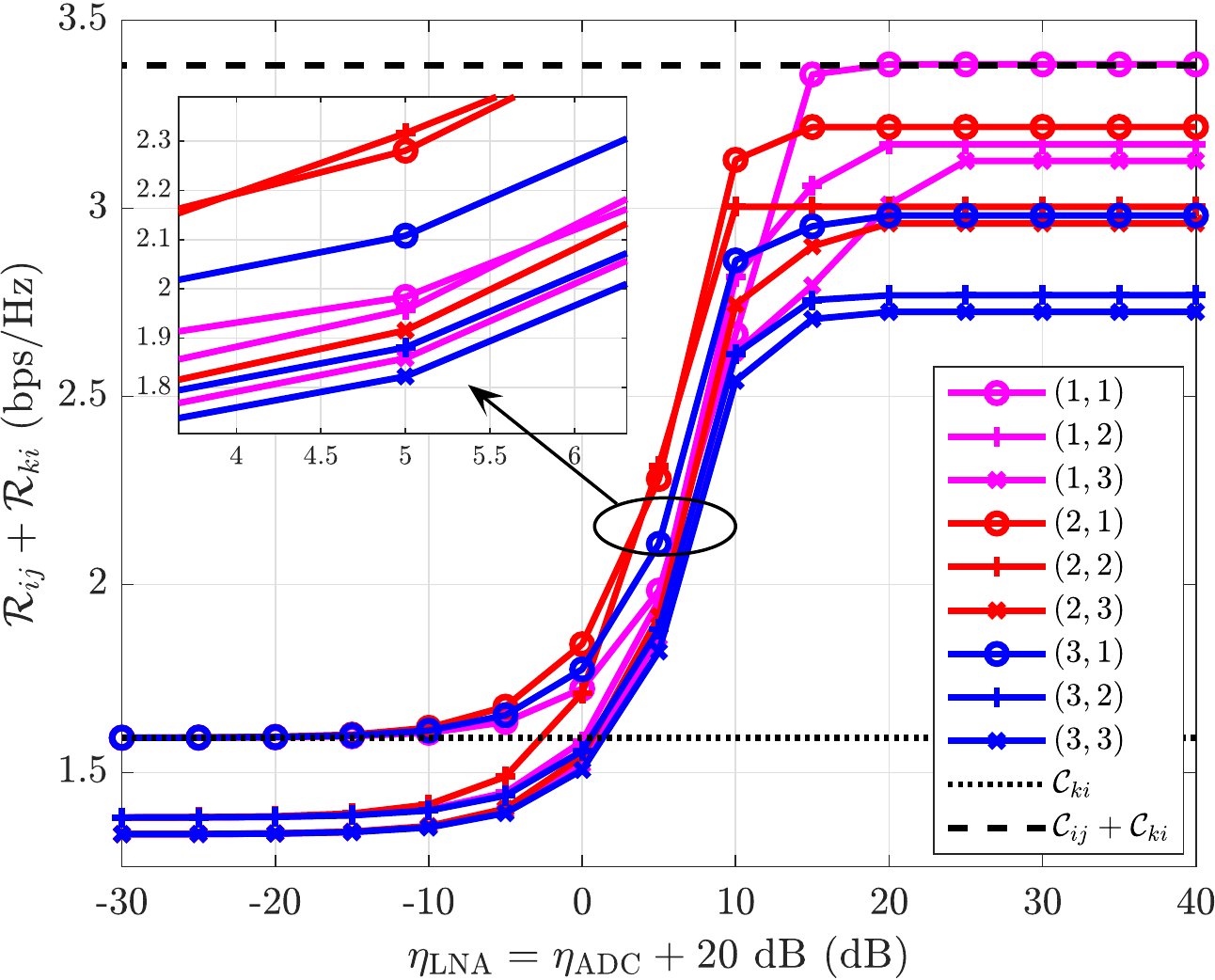}
		\label{fig:results-cand-eta-b}}
	\caption{Spectral efficiency as a function of $\etalna = \etaadc + 20$ dB for various candidates from $\Tij$ and $\Tki$, where $\Kij = \Kki = 3$, $\kappa = 10$ dB, $\adcbits = 12$ bits, $\snr_{ij} = \snr_{ki} = -10$ dB.}
	\label{fig:results-cand-eta}
\end{figure*}

We use \figref{fig:results-cand-eta} to demonstrate the essence behind our design's approach to using the candidate sets $\Tij$ and $\Tki$ to further maximize the spectral efficiency delivered from $i$ to $j$ while meeting our \lna and \adc constraints.
In \figref{fig:results-cand-eta-a}, we plot $\Rij$ as a function of $\etalna = \etaadc + 20$ dB for each of the nine candidates when $\Kij = \Kki = 3$, where $(t,r)$ denotes using the $t$-th candidate from $\Tij$ and the $r$-th candidate from $\Tki$.
When $\etalna = \etaadc + 20$ dB is sufficiently large, we can see that transmission from $i$ to $j$ is effectively unconstrained, meaning the capacity-achieving strategy for a given transmit candidate implicitly satisfies the \lna and \adc constraints.
In such a case, choosing the first candidate from $\Tij$ (i.e., candidate $(1,1)$, $(1,2)$, or $(1,3)$) is capacity-optimal as denoted by $\Cij$ coinciding with $\Rij$ at high $\etalna = \etaadc + 20$ dB.
As $\etalna = \etaadc + 20$ dB decreases, different candidates begin to fall off in their achievable spectral efficiency $\Rij$ to meet the constraints.
Some candidates are more robust as $\etalna = \etaadc + 20$ dB decreases, allowing them to achieve higher $\Rij$ versus other candidates.
For a given set of conditions, including $\etalna = \etaadc + 20$ dB, our design effectively chooses the optimal candidate.
Depending on the channel realizations, some candidates may be significantly more advantageous than others.
In this modest example, \figref{fig:results-cand-eta-a} shows that, at $\etalna = 5$ dB ($\etaadc = -15$ dB), choosing candidate $(2,2)$ over $(1,1)$ results in more than doubling $\Rij$.

\figref{fig:results-cand-eta-b} extends these results to the sum spectral efficiency .
Still, at $\etalna = 5$ dB, candidate $(2,2)$ maximizes $\Rij + \Rki$, though the gap between it and candidate $(2,1)$ has significantly shrunk and are nearly overlapping.
This suggests that the gains achieved by candidate $(2,2)$ over $(2,1)$ in $\Rij$ are approximately equal to the sacrifice in $\Rki$ by choosing candidate $2$ instead of $1$ from $\Tki$.
Noticing that candidate pairs converging to $\Cki$ suggests that candidate $1$ from $\Tki$ is in fact optimal in maximizing $\Rki$ alone.
Again, this highlights the importance of a system being aware when it is in regimes where it is better off operating in half-duplex rather than full-duplex.

---
}


\begin{figure}[!t]
	\centering
	\includegraphics[width=0.47\linewidth,height=0.4\textheight,keepaspectratio]{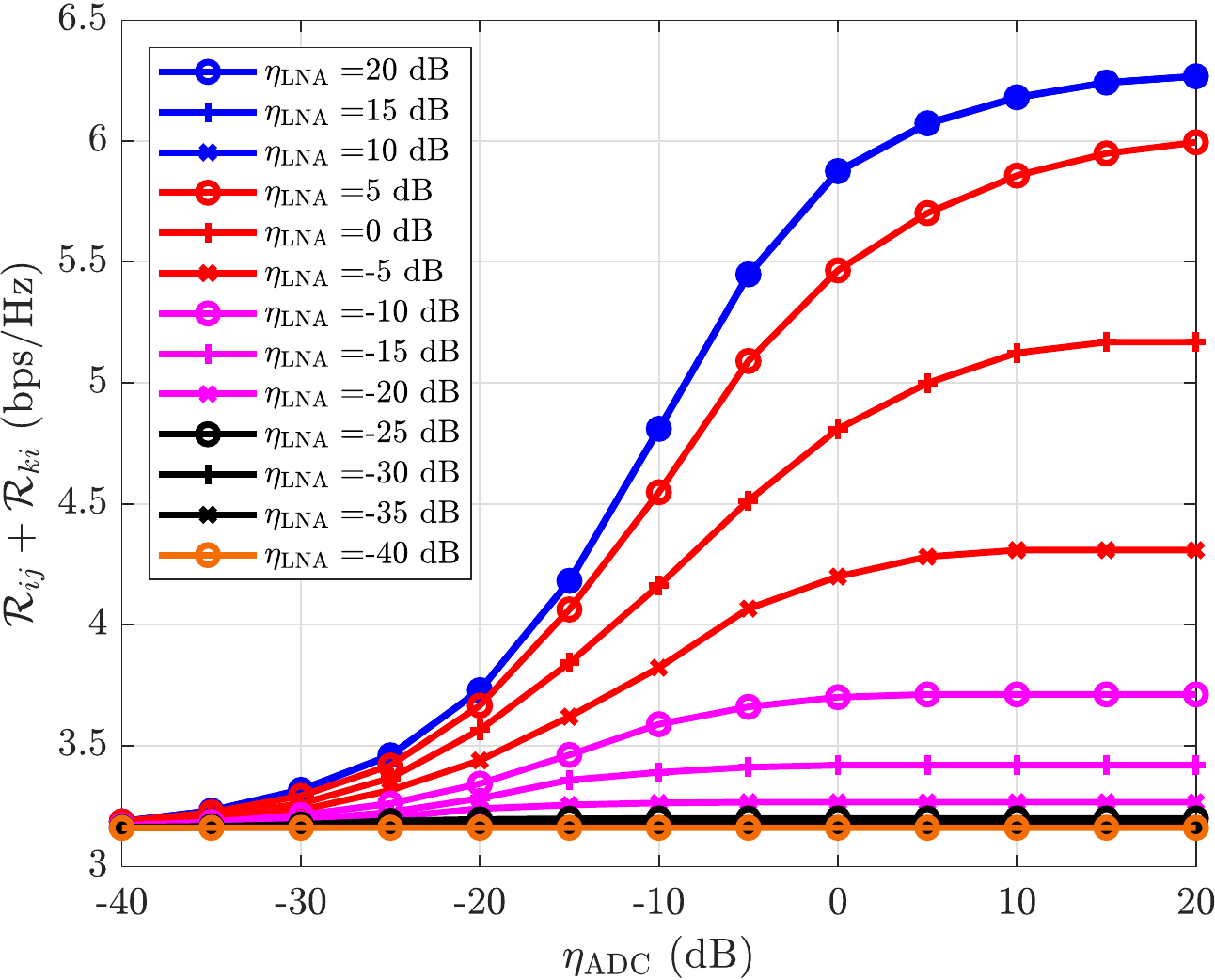}
	\caption{Sum spectral efficiency as a function of $\etaadc$ for various $\etalna$, where $\kappa = 10$ dB, $K_{ij} = K_{ki} = 1$, $\snr_{ij} = \snr_{ki} = -10$ dB, and $b = 12$ bits. As $\etalna$ and $\etaadc$ are relaxed, the system can achieve a greater sum spectral efficiency since transmit performance is less constrained by a limited receive dynamic range.}
	\label{fig:results-se-eta}
\end{figure}

In \figref{fig:results-se-eta}, we evaluate the sum spectral efficiency $\Rij + \Rki$ for various selections of $\etalna$ and $\etaadc$.
To better illustrate this, we let $\Kij = \Kki = 1$, placing the sole responsibility of preventing saturation on $\prebb\node{i}$ and making $\Rki$ approximately constant ($\approx 3.2$ bps/Hz) across all $\etalna$, $\etaadc$ (since $\adcbits = 12$ bits).
Intuitively, as $\etalna$ and $\etaadc$ increase, optimizing transmission from $i$ becomes more relaxed, allowing for higher $\Rij$.
For a given choice of $\etalna$, we can see that at increasing $\etaadc$ beyond some point has little to no effect on $\Rij$.
This can be attributed to the fact that \textbf{the \lna constraint begins to supersede the \adc constraint} at these points (recall \thmref{thm:adc-redundant-by-lna}).
As \etalna increases, the point at which \etaadc plays no role also increases.
Furthermore, beyond a certain $\etalna$ (e.g., $\etalna \geq 10$ dB), we can see that the \lna constraint becomes immaterial, suggesting that the precoding power constraint implicitly satisfies the \lna constraint (recall \thmref{thm:lna-redundant-by-power}).

\begin{figure*}[!t]
	\centering
	\subfloat[Receive link spectral efficiency.]{\includegraphics[width=0.47\linewidth,height=0.4\textheight,keepaspectratio]{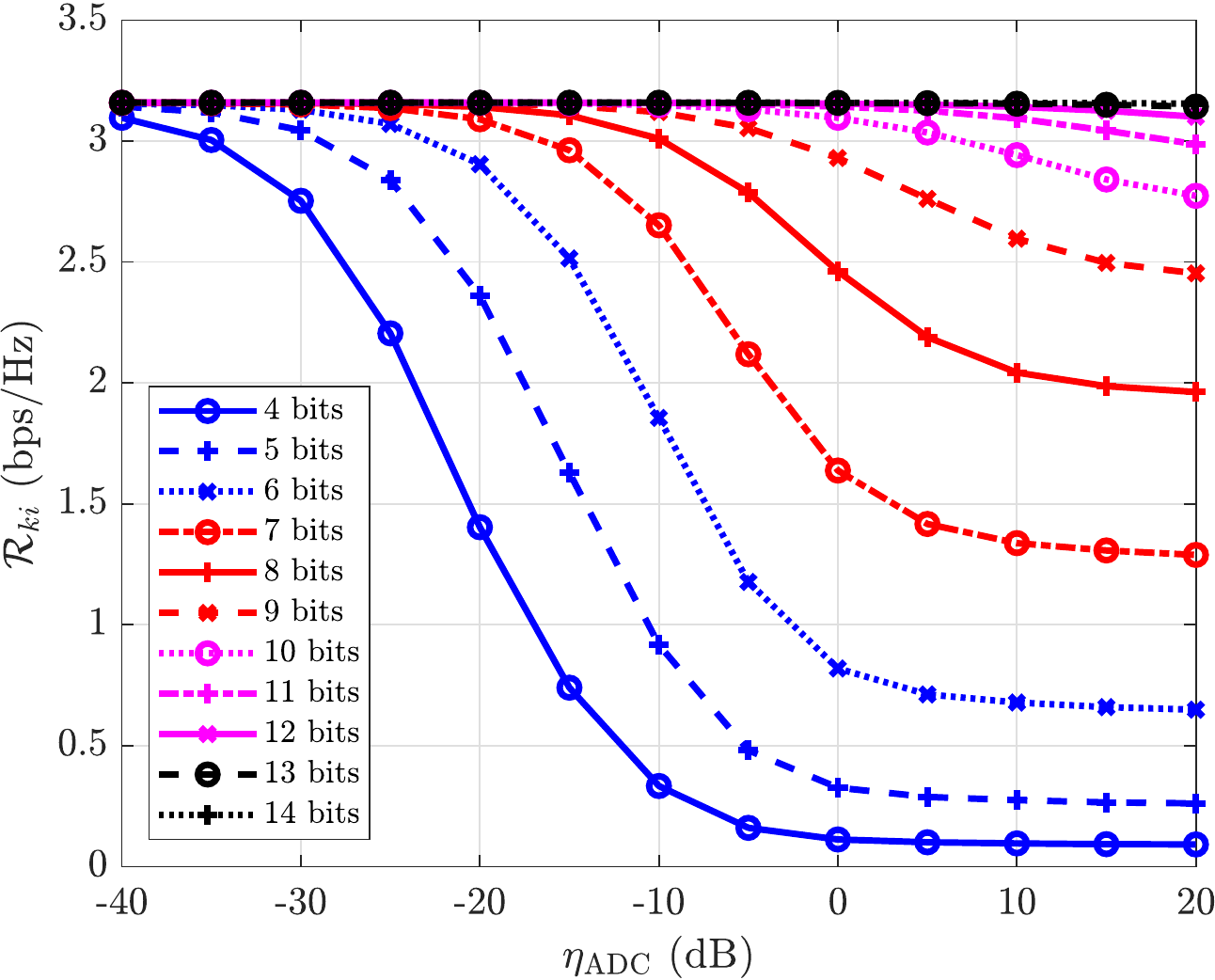}
		\label{fig:results-se-eta-bits-a}}
	\quad
	\subfloat[Sum spectral efficiency.]{\includegraphics[width=0.47\linewidth,height=0.4\textheight,keepaspectratio]{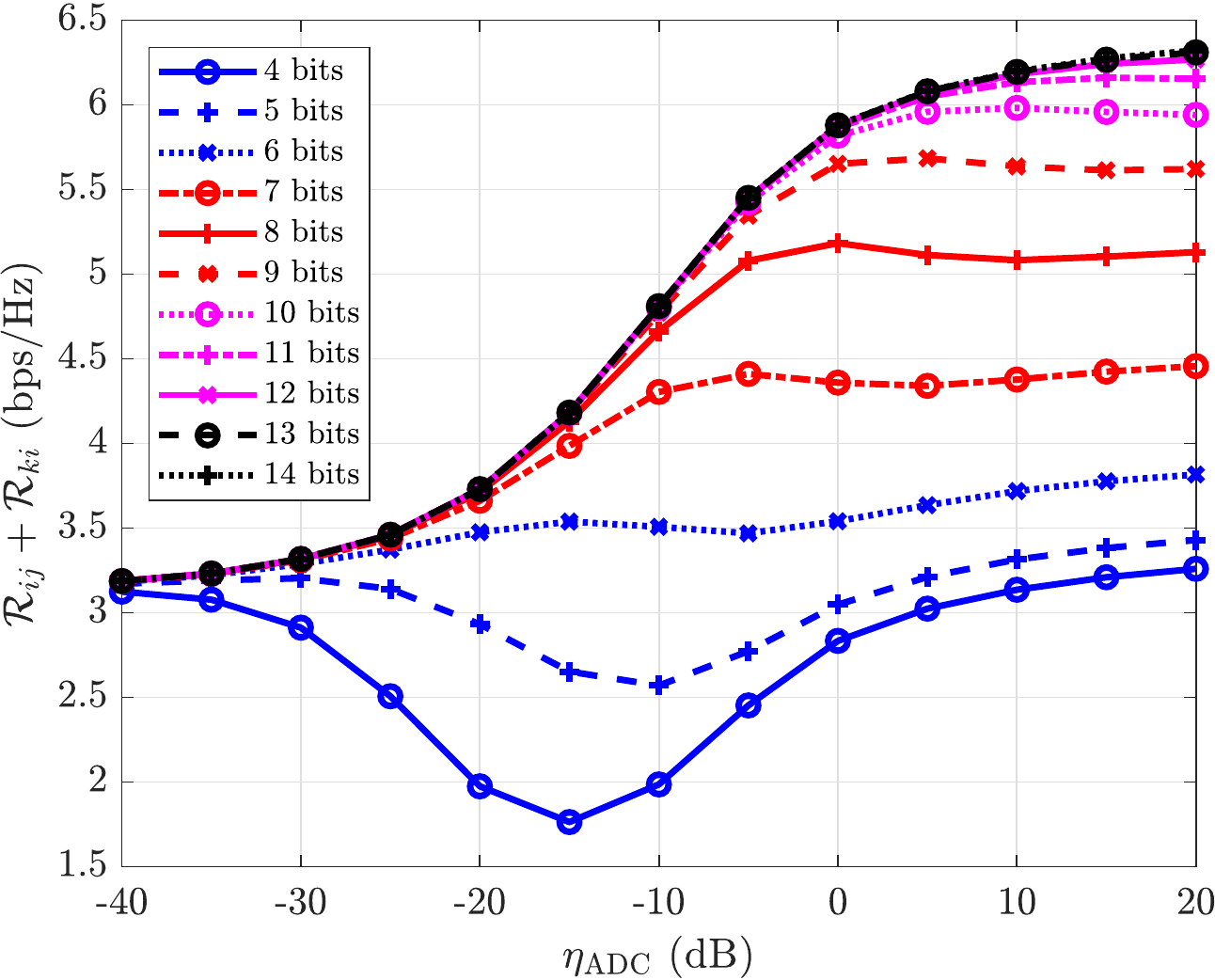}
		\label{fig:results-se-eta-bits-b}}
	\caption{Spectral efficiency as a function of $\etaadc$ for various \adc resolutions, where $\kappa = 10$ dB, $K_{ij} = K_{ki} = 1$, $\snr_{ij} = \snr_{ki} = -10$ dB, and $\etalna = 20$ dB. Higher-resolution \adcs are more robust to self-interference whereas lower-resolution \adcs saturate if $\etaadc$ is not properly chosen. The tradeoff associated with constraining the transmit link performance and preventing \adc saturation is not so obvious in terms of sum spectral efficiency.}
	\label{fig:results-se-eta-bits}
\end{figure*}

Now, we examine the importance of $\etaadc$ for various \adc resolutions.
A key motivator for this work is the fact that self-interference can increase quantization noise, degrading the effective \gls{snr} of a desired signal out of the \adc.
\adcs having greater resolution have a higher dynamic range, allowing them to quantize signal-plus-interference-plus-noise without suffering from \adc saturation as severely as lower-resolution \adcs.
This can be seen in \figref{fig:results-se-eta-bits-a}, where we have evaluated $\Rki$ as a function of $\etaadc$ for various \adc resolutions, taking $\etalna = 20$ dB and $\Kij = \Kki = 1$ to reduce their impacts on interpreting these results.
Higher-resolution \adcs are practically invariant across \etaadc, allowing them to achieve approximately the same $\Rki$ regardless of the relative self-interference power at the \adcs.
As the resolution decreases, we can see that quantization noise begins to take its toll on the spectral efficiency \Rki, where it eventually plateaus beyond a certain $\etaadc$ as the other constraints take effect.
This highlights that an appropriate choice of $\etaadc$ is intimately connected with the resolution of the \adcs and further justifies the motivation for this work: \textbf{under limited \adc resolution, the need to limit the self-interference power reaching the \adcs is critical}.

Interesting things happen in terms of the sum spectral efficiency $\Rij + \Rki$, with varying $\etaadc$ and \adc resolutions, as depicted in \figref{fig:results-se-eta-bits-b}.
As discussed, high-resolution \adcs are relatively invariant to \etaadc, and therefore, changes in sum spectral efficiency can be attributed almost exclusively to changes in $\Rij$ as a function of \etaadc.
With low-resolution \adcs (e.g., $b=4,5$ bits), we see that $\Rij + \Rki$ initially decreases sharply as $\etaadc$ increases.
This is due to the falloff that we saw in \figref{fig:results-se-eta-bits-a}.
At very low (strict) $\etaadc$, $\Rij$ is also very low, meaning the sharp falloff in \Rki drastically degrades the sum spectral efficiency.
As \etaadc is increased (e.g., $\etaadc = -10$ dB), $\Rij$ also increases while $\Rki$ begins to plateau.
Recall that $\Rij$ is invariant to the \adc resolution at $i$.
As \etaadc further increases (e.g., $\etaadc = 0$ dB), $\Rij$ further increases and more rapidly so as its \adc saturation requirements become even more relaxed whereas $\Rki$ further plateaus.
Finally, as $\etaadc$ increases further (e.g., $\etaadc \geq 10$ dB), $\Rki$ remains plateaued and $\Rij$ begins to also plateau as it sees less gain in $\Rij$ as changes in $\etaadc$ hold less meaning, given the presence of $\etalna = 20$ dB and the precoding power constraint.
For \adcs falling between very high resolutions and very low resolutions, the behavior can be explained in a similar fashion by this intertwining of $\Rij$ and $\Rki$, both of which begin to saturate beyond a certain $\etaadc$.

\figref{fig:results-se-eta-bits-b} highlights an important fact:
\textbf{an appropriate design is necessary to make full-duplex operation worthwhile over half-duplex}.
This is evidenced by the fact that low-resolution \adcs (e.g., $\adcbits = 4,5$ bits) demand such significant self-interference mitigation that the sacrifice made on the transmit link is not worthwhile.
Furthermore, we can see that the ($\Rij+\Rki$)-optimal degree of self-interference power permitted at the \adcs varies with resolution.
At these optimal choices for $\etaadc$, introducing a modest amount of quantization noise to improve transmit link performance balances $\Rij$ and $\Rki$ such that their sum is maximized.


\begin{figure}[!t]
	\centering
	\includegraphics[width=0.47\linewidth,height=0.4\textheight,keepaspectratio]{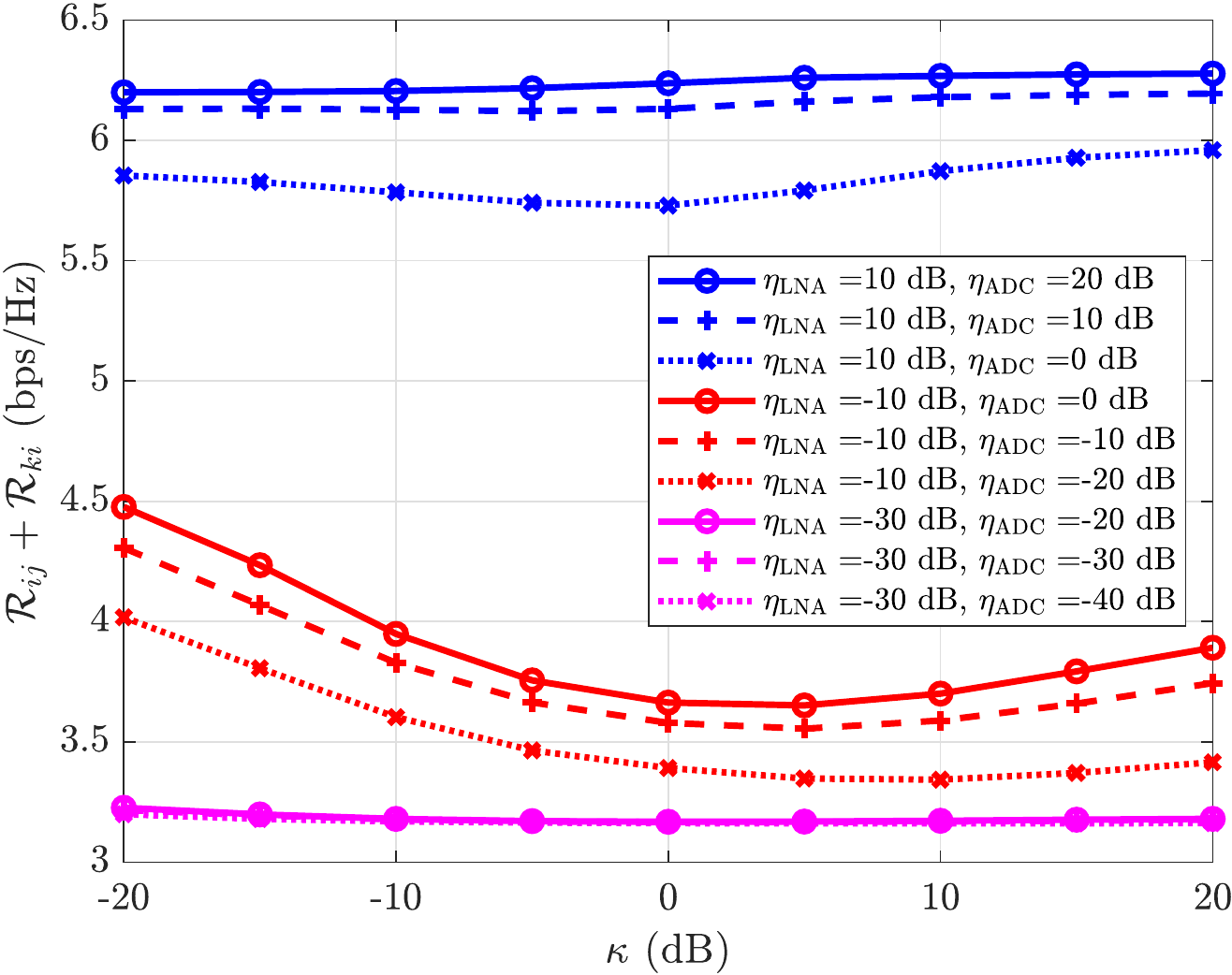}
	\caption{Sum spectral efficiency as a function of $\kappa$ for various $\etalna$ and $\etaadc$, where $K_{ij} = K_{ki} = 1$, $\snr_{ij} = \snr_{ki} = -10$ dB, and $b = 12$ bits.}
	\label{fig:results-se-kappa}
\end{figure}

Finally, for the sake of completeness and to better understand the role $\kappa$ (the self-interference channel's Rician factor) from \eqref{eq:si-channel-rice} plays in our design, we have included \figref{fig:results-se-kappa}.
We have evaluated the sum spectral efficiency for various pairs of \etalna and \etaadc as a function of $\kappa$, fixing all other variables.
Under relaxed conditions (i.e., high \etalna, \etaadc), avoiding the self-interference channel becomes less of a priority, rendering $\kappa$ less impactful.
Under stringent conditions (i.e., low \etalna, \etaadc), we can also see that $\kappa$ does not play much of a role.
This can be attributed to the fact that satisfying these strict \lna and \adc constraints is done so largely by power control of $\prebb\node{i}$ rather than through steering strategies.
This renders the underlying structure of $\channel_{ii}$, and thus $\kappa$, less of a factor.
In between, however, we see that modest choices of \etalna and \etaadc makes the role of $\kappa$ more significant.
When $\kappa$ is very low, the self-interference channel is comprised primarily of far-field reflections. 
This leads to a self-interference channel that is spatially sparse, meaning avoiding $\channel_{ii}$ becomes easier with highly directional \gls{dft} beams.
This can be similarly stated that the inherent isolation between the rays of $\channel_{ij}$ and $\channel_{ii}$ and of $\channel_{ki}$ and $\channel_{ii}$ leads to more easily avoiding pushing self-interference onto the receiver of our full-duplex device.
Similar behavior happens when $\kappa$ is high, though the spatial sparsity of the self-interference channel stems from the near-field channel structure produced by our vertically-separated horizontal uniform linear arrays at the full-duplex device.
Between, when $\kappa$ approaches zero, the two spatially sparse channel components---the far-field portion and the near-field portion---mix relatively evenly, which leads to a self-interference channel that is less spatially sparse, making it more difficult to avoid pushing energy into with highly directional \gls{dft} beams.
This leads to a relatively lower $\Rij$ and, thus, lower sum spectral efficiency.


\section{Conclusion} \label{sec:conclusion}

We have presented a hybrid beamforming design for \mmwave full-duplex that holistically accounts for a number of practical considerations including codebook-based analog beamforming and beam alignment, a desirably low number of \rf chains, and the need to prevent receiver-side saturation at the full-duplex device.
Core to our design is its focus on limiting the self-interference power reaching each antenna and each \rf chain, to prevent pushing \lnas beyond their linear region and avoid flooding a desired signal with quantization noise at the \adcs.
Our design utilizes sets of candidate analog beamformers to improve its flexibility in mitigating self-interference while maintaining service and to accommodate codebook-based analog beamforming.
Numerical results have highlighted the costs and limitations associated with preventing receiver-side saturation, which can be used in \mmwave full-duplex system analyses when choosing components, constructing a \mmwave full-duplex transceiver, and determining what levels of self-interference mitigation should be aimed for.
Useful future work would explore the integration of full-duplex \mmwave transceivers into cellular standards, implementation of beamforming-based self-interference mitigation, and characterization of \mmwave self-interference channels.







\bibliographystyle{bibtex/IEEEtran}
\bibliography{bibtex/IEEEabrv,refs_v2}

\begin{thebibliography}{10}
\providecommand{\url}[1]{#1}
\csname url@samestyle\endcsname
\providecommand{\newblock}{\relax}
\providecommand{\bibinfo}[2]{#2}
\providecommand{\BIBentrySTDinterwordspacing}{\spaceskip=0pt\relax}
\providecommand{\BIBentryALTinterwordstretchfactor}{4}
\providecommand{\BIBentryALTinterwordspacing}{\spaceskip=\fontdimen2\font plus
\BIBentryALTinterwordstretchfactor\fontdimen3\font minus
  \fontdimen4\font\relax}
\providecommand{\BIBforeignlanguage}[2]{{%
\expandafter\ifx\csname l@#1\endcsname\relax
\typeout{** WARNING: IEEEtran.bst: No hyphenation pattern has been}%
\typeout{** loaded for the language `#1'. Using the pattern for}%
\typeout{** the default language instead.}%
\else
\language=\csname l@#1\endcsname
\fi
#2}}
\providecommand{\BIBdecl}{\relax}
\BIBdecl

\bibitem{sabharwal_-band_2014}
A.~Sabharwal \emph{\textit{et~al.}}, ``In-band full-duplex wireless: Challenges
  and opportunities,'' \emph{{IEEE} JSAC}, vol.~32, no.~9, pp. 1637--1652, Sep.
  2014.

\bibitem{xia_2017}
Z.~{Xiao}, P.~{Xia}, and X.~{Xia}, ``Full-duplex millimeter-wave
  communication,'' \emph{{IEEE} Wireless Commun.}, vol.~24, no.~6, pp.
  136--143, 2017.

\bibitem{roberts_wcm}
I.~P. Roberts, J.~G. Andrews, H.~B. Jain, and S.~Vishwanath, ``Millimeter wave
  full-duplex radios: New challenges and techniques,'' \emph{{IEEE} Wireless
  Commun.}, Feb. 2021.

\bibitem{Huberman_Le-Ngoc_2015}
S.~Huberman and T.~Le-Ngoc, ``{MIMO} full-duplex precoding: A joint beamforming
  and self-interference cancellation structure,'' \emph{{IEEE} Trans. Wireless
  Commun.}, vol.~14, no.~4, pp. 2205--2217, Apr. 2015.

\bibitem{everett_softnull_2016}
E.~Everett, C.~Shepard, L.~Zhong, and A.~Sabharwal, ``{SoftNull}: Many-antenna
  full-duplex wireless via digital beamforming,'' \emph{{IEEE} Trans. Wireless
  Commun.}, vol.~15, no.~12, pp. 8077--8092, Dec. 2016.

\bibitem{Alexandropoulos_Duarte_2017}
G.~C. Alexandropoulos and M.~Duarte, ``Joint design of multi-tap analog
  cancellation and digital beamforming for reduced complexity full duplex
  {MIMO} systems,'' in \emph{Proc. {IEEE} Intl. Conf. Commun.}, May 2017, pp.
  1--7.

\bibitem{dinc_60_2016}
T.~Dinc, A.~Chakrabarti, and H.~Krishnaswamy, ``A 60 {GHz} {CMOS} full-duplex
  transceiver and link with polarization-based antenna and {RF} cancellation,''
  \emph{{IEEE} J. Solid-State Circuits}, vol.~51, no.~5, pp. 1125--1140, May
  2016.

\bibitem{dinc_2017}
T.~Dinc \emph{\textit{et~al.}}, ``Synchronized conductivity modulation to
  realize broadband lossless magnetic-free non-reciprocity,'' \emph{Nature
  Commun.}, vol.~8, no.~11, pp. 1--9, Oct 2017.

\bibitem{singh_2020_acm}
V.~Singh \emph{\textit{et~al.}}, ``Millimeter-wave full duplex radios,'' in
  \emph{Proc. {ACM} Intl. Conf. Mob. Comput. and Netw. (MobiCom)}, 2020.

\bibitem{satyanarayana_hybrid_2019}
K.~Satyanarayana, M.~El-Hajjar, P.~Kuo, A.~Mourad, and L.~Hanzo, ``Hybrid
  beamforming design for full-duplex millimeter wave communication,''
  \emph{{IEEE} Trans. Veh. Technol.}, vol.~68, no.~2, pp. 1394--1404, Feb.
  2019.

\bibitem{liu_beamforming_2016}
X.~Liu \emph{\textit{et~al.}}, ``Beamforming based full-duplex for
  millimeter-wave communication,'' \emph{Sensors}, vol.~16, no.~7, p. 1130,
  Jul. 2016.

\bibitem{lopez_prelcic_2019_analog}
R.~L{\'o}pez-Valcarce and N.~Gonz{\'a}lez-Prelcic, ``Analog beamforming for
  full-duplex millimeter wave communication,'' in \emph{Proc. Intl. Symp.
  Wireless Commun. Syst.}, Aug. 2019, pp. 687--691.

\bibitem{lopez-valcarce_beamformer_2019}
------, ``Beamformer design for full-duplex amplify-and-forward millimeter wave
  relays,'' in \emph{Proc. Intl. Symp. Wireless Commun. Syst.}, Aug. 2019, pp.
  86--90.

\bibitem{prelcic_2019_hybrid}
J.~{Palacios}, J.~{Rodr{\'i}guez-Fern{\'a}ndez}, and N.~{Gonz{\'a}lez-Prelcic},
  ``Hybrid precoding and combining for full-duplex millimeter wave
  communication,'' in \emph{Proc. {IEEE} Global Commun. Conf.}, 2019, pp. 1--6.

\bibitem{roberts_2019_bfc}
I.~P. {Roberts} and S.~{Vishwanath}, ``Beamforming cancellation design for
  millimeter-wave full-duplex,'' in \emph{Proc. {IEEE} Global Commun. Conf.},
  2019.

\bibitem{cai_robust_2019}
Y.~Cai, Y.~Xu, Q.~Shi, B.~Champagne, and L.~Hanzo, ``Robust joint hybrid
  transceiver design for millimeter wave full-duplex {MIMO} relay systems,''
  \emph{{IEEE} Trans. Wireless Commun.}, vol.~18, no.~2, pp. 1199--1215, Feb.
  2019.

\bibitem{roberts_2020_fsbfc}
I.~P. Roberts, H.~B. Jain, and S.~Vishwanath, ``Frequency-selective beamforming
  cancellation design for millimeter-wave full-duplex,'' in \emph{Proc. {IEEE}
  Intl. Conf. Commun.}, 2020, pp. 1--6.

\bibitem{roberts_equipping_2020}
------, ``Equipping millimeter-wave full-duplex with analog self-interference
  cancellation,'' in \emph{Proc. {IEEE} Intl. Conf. Commun. Wkshp.}, Jun. 2020.

\bibitem{zhu_uav_joint_2020}
L.~{Zhu} \emph{\textit{et~al.}}, ``Millimeter-wave full-duplex {UAV} relay:
  Joint positioning, beamforming, and power control,'' \emph{{IEEE} JSAC},
  vol.~38, no.~9, pp. 2057--2073, 2020.

\bibitem{heath_overview_2016}
R.~W. Heath, N.~Gonz{\'a}lez-Prelcic, S.~Rangan, W.~Roh, and A.~M. Sayeed, ``An
  overview of signal processing techniques for millimeter wave {MIMO}
  systems,'' \emph{{IEEE} J. Sel. Topics Signal Process.}, vol.~10, no.~3, pp.
  436--453, Apr. 2016.

\bibitem{rf_gu_2005}
Q.~Gu, \emph{{RF} System Design of Transceivers for Wireless
  Communications}.\hskip 1em plus 0.5em minus 0.4em\relax Springer, 2005.

\bibitem{Day_Margetts_Bliss_Schniter_2012}
B.~P. Day, A.~R. Margetts, D.~W. Bliss, and P.~Schniter, ``Full-duplex {MIMO}
  relaying: Achievable rates under limited dynamic range,'' \emph{{IEEE} JSAC},
  vol.~30, no.~8, pp. 1541--1553, Sep 2012.

\bibitem{heath_lozano}
R.~W. Heath~Jr. and A.~Lozano, \emph{Foundations of {MIMO}
  Communication}.\hskip 1em plus 0.5em minus 0.4em\relax Cambridge University
  Press, 2018.

\bibitem{cvx}
M.~Grant and S.~Boyd, ``{CVX}: Software for disciplined convex programming,
  version 2.1,'' \url{http://cvxr.com/cvx}, Mar. 2014.

\bibitem{spherical_2005}
J.-S. Jiang and M.~A. Ingram, ``Spherical-wave model for short-range {MIMO},''
  \emph{{IEEE} Trans. Commun.}, vol.~53, no.~9, pp. 1534--1541, 2005.

\end{thebibliography}

\section*{Acknowledgments}

This work was supported by the National Science Foundation Graduate Research Fellowship Program (Grant No.~DGE-1610403).
Any opinions, findings, and conclusions or recommendations expressed in this material are those of the author(s) and do not necessarily reflect the views of the National Science Foundation.



\end{document}